\documentclass[twocolumn,aps,amssymb,noshowpacs,superscriptaddress,nofootinbib]{revtex4}
\usepackage{amsmath,amssymb}
\usepackage{graphicx}
\usepackage{verbatim}
\usepackage{amsfonts}
\usepackage{dcolumn}
\usepackage{bm}
\usepackage[section]{placeins}
\usepackage{color}
\usepackage{mathrsfs}
\usepackage[colorlinks,bookmarks=true,citecolor=blue,linkcolor=red,urlcolor=blue]{hyperref}
\usepackage{afterpage}
\usepackage{ulem}
\usepackage{url} 
\usepackage[T1]{fontenc}
\usepackage{lmodern}
\usepackage[utf8]{inputenc} 
\usepackage{chemformula}
\usepackage{mathrsfs}
\usepackage{ntheorem}

\newtheorem{theorem}{Theorem}
\newtheorem{lemma}{Lemma}

\newtheorem{define}{Definition}

\newtheorem{corollary}{Corollary}

\usepackage{longtable}
\usepackage{multirow}

\newcommand{\beq}{\begin{eqnarray} }
\newcommand{\eeq}{\end{eqnarray} }
\newcommand{\Beq}{\begin{eqnarray*} }
\newcommand{\Eeq}{\end{eqnarray*} }

\newcommand{\RNum}[1]{\uppercase\expandafter{\romannumeral #1\relax}}

\newcommand{\beginsupplement}{%
        \setcounter{table}{0}
        \renewcommand{\thetable}{S\arabic{table}}%
        \setcounter{figure}{0}
        \renewcommand{\thefigure}{S\arabic{figure}}%
        \setcounter{section}{0}
        \renewcommand{\thesection}{S\arabic{section}}%
        \setcounter{equation}{0}
        \renewcommand{\theequation}{S\arabic{equation}}%
     }

\begin{document}
\draft

\title{Constructions and Applications of Irreducible Representations of Spin-Space Groups}
\author{Ziyin Song}
 \thanks{These authors contributed equally to this study.}
 \affiliation{Beijing National Laboratory for Condensed Matter Physics, and Institute of Physics, Chinese Academy of Sciences, Beijing 100190, China}
 \affiliation{University of Chinese Academy of Sciences, Beijing 100049, China}
 \author{A. Z. Yang}
 \thanks{These authors contributed equally to this study.}
\affiliation{School of Physics and Beijing Key Laboratory of Opto-electronic Functional Materials and Micro-nano Devices, Renmin University of China, Beijing, 100872, China}
\affiliation{Key Laboratory of Quantum State Construction and Manipulation (Ministry of Education), Renmin University of China, Beijing, 100872, China}
 \author{Yi Jiang}
 \affiliation{Beijing National Laboratory for Condensed Matter Physics, and Institute of Physics, Chinese Academy of Sciences, Beijing 100190, China} 
 \affiliation{University of Chinese Academy of Sciences, Beijing 100049, China}
 \affiliation{Donostia International Physics Center (DIPC), Paseo Manuel de Lardizábal. 20018, San Sebastián, Spain}
 \author{Zhong Fang}
 \affiliation{Beijing National Laboratory for Condensed Matter Physics, and Institute of Physics, Chinese Academy of Sciences, Beijing 100190, China}
 \affiliation{Songshan Lake Materials Laboratory, Dongguan, Guangdong 523808, China}
 \author{Jian Yang} 
 \affiliation{Beijing National Laboratory for Condensed Matter Physics, and Institute of Physics, Chinese Academy of Sciences, Beijing 100190, China}
  \author{Chen Fang}
 \affiliation{Beijing National Laboratory for Condensed Matter Physics, and Institute of Physics, Chinese Academy of Sciences, Beijing 100190, China}
 \affiliation{Songshan Lake Materials Laboratory, Dongguan, Guangdong 523808, China}
 \affiliation{Kavli Institute for Theoretical Sciences, Chinese Academy of Sciences, Beijing 100190, China}
 \author{Hongming Weng}
  \email{hmweng@iphy.ac.cn}
 \affiliation{Beijing National Laboratory for Condensed Matter Physics, and Institute of Physics, Chinese Academy of Sciences, Beijing 100190, China}
 \author{Zheng-Xin Liu}
 \email{liuzxphys@ruc.edu.cn}
\affiliation{School of Physics and Beijing Key Laboratory of Opto-electronic Functional Materials and Micro-nano Devices, Renmin University of China, Beijing, 100872, China}
\affiliation{Key Laboratory of Quantum State Construction and Manipulation (Ministry of Education), Renmin University of China, Beijing, 100872, China}

\date{\today}
\begin{abstract}
Spin-space groups (SSGs), including the traditional space groups (SGs) and magnetic space groups (MSGs) as subsets, describe the complete symmetries of magnetic materials with weak spin-orbit coupling (SOC). In the present work, we systematically study the irreducible representations (irreps) of SSGs by focusing on the projective irreps of the little co-group $L(k)$ of any momentum point $\pmb k$. We analysis the factor systems of $L(k)$, and then reduce the projective regular representation of $L(k)$ into direct sum of irreps using the Hamiltonian approach. Especially, for collinear SSGs which contain continuous spin rotation operations, we adopt discrete subgroups to effectively capture their characteristics. Furthermore, we apply the representation theory of SSGs to study the 
band structure of electrons and magnons in magnetic materials.
After identifying the SSG symmetry group, we extract relevant irreps and determine the $k\cdot p$ models. As an example, we illustrate how our approach works for the material \ch{Mn3Sn}.  
Degeneracies facilitated by SSG symmetry are observed, underscoring the effectiveness of application in material analysis. The SSG recognition and representation code is uploaded to \href{https://github.com/zine-phy/MOM2SSG}{GitHub}, the information of irreps of all SSGs is also available in the \href{https://cmpdc.iphy.ac.cn/ssg/}{online Database}. Our work provides a practical toolkit for exploring the intricate symmetries of magnetic materials and paves the way for future advances in materials science.

\end{abstract}

\maketitle

\section{Introduction}

Symmetries play a essential role in understanding the physical laws governing our universe. This principle holds true in condensed matter physics, where symmetries significantly facilitate the investigation of the properties and behaviors of crystalline materials.
Symmetry analysis is crucial for classifying phases of matter, for predicting physical properties of matters, and for understanding their microscopic structures as well. Different systems require specific groups for a precise and complete description of their symmetry. For non-magnetic systems, one can use the 230 SGs\cite{hahn1983international, bradley2010} to describe the microscopic symmetries of various crystals.   When considering magnetic orders on the lattice, generally we need to include anti-unitary symmetry elements including the time-reversal operation. This gives rise to totally 1651 MSGs including the SGs as a special type \cite{Bradley1968, Lifshitz2004, bradley2010, Litvin2013MagneticGT, gallego2016magndataI, gonzalez2021extension, LiuYaomsgcorep2023}. The extensive study of these groups greatly enriched our knowledge for the physical properties and their applications of magnetic materials. 

The concept of SSG\cite{Brinkman1966, Litvin1974} was proposed to depict magnetic systems with weak SOC, in complement to SGs and MSGs. Recently, the investigation of SSGs has been attracting increasing attentions, from the complete enumeration of the group structures \cite{xiao2023spin, ren2023enumeration,jiang2023enumeration}, the investigation of new quasi-particle types\cite{YangLiuFang2021, pjguo2021prl, pjguo2022, Liu2022prx, liu2022chiral}, to the exploration of alter-magnetic materials\cite{smejkal2021altermagnetism, TJungwirth2022, mazin2022altermagnetism} and non-collinear/non-coplanar antiferromagnetism\cite{Chen2014AHE, Suzuki2017AHE}. 
As SGs and MSGs are subsets of SSGs, SSGs provide a more comprehensive perspective of space-time symmetries in condensed matter physics.  

The representations (reps) of groups play a critical role in applying symmetry rules to analyze physical properties of matters, including the classification of gapped topological phases\cite{Peng2022topo, Elcoro2021, Fang2012topo, Po2017, Bradlyn2017}, the $k\cdot p$ theory of quasi-particles \cite{Willatzen2009, LuttingerKP, Gresch2018, Jiang_2021, Yang2020so4}, and the determination of crystal and magnetic structures via scattering experiments\cite{rodriguez2001fullprof, wills2000new}.  The applications of group theory inherently demands the information of all inequivalent irreps. These information of SGs and MSGs are well-established and accessible through the Bilbao Crystallographic Server\cite{Aroyo1, Aroyo2, Aroyo3}. However a complete information of irreps of SSGs is still lacking.

Previous works on the representation theory of SSGs have encompassed the construction of reps for certain spin point groups\cite{schiff2023spin,YangLiuFang2021} or an introductory information of the irreps of SSGs\cite{ren2023enumeration, xiao2023spin, jiang2023enumeration}. Therefore, a 
comprehensive study of all irreps of SSGs, especially the projective irreps of little co-groups at all high-symmetry points or high-symmetry lines, is essential for their broader applications in solid-state physics. In the present work, we present a high efficient algebraic approach to calculate all irreps of the little co-groups at high symmetry $k$ points, and incorporate the data into our database (\url{https://cmpdc.iphy.ac.cn/ssg/}). Then we outline a methodology for applying SSGs to describe the symmetries and physical properties (such as the band structure for the electrons or magnons) of magnetic materials and illustrate the approach via a concrete example \ch{Mn3Sn}. 

The paper is organized as follows: In Sec. \ref{anti-unitary representations}, the representation theory of anti-unitary groups with a non-trivial factor system is discussed. 
In Sec. \ref{little group}, we introduce the structure of SSGs and their little co-groups at high symmetry momentum points.  Sec. \ref{rep of SSG} applies this representation theory to SSGs, employing the `Hamiltonian method' for reduction to obtain all irreps. 
In Sec. \ref{ultilities of SSG representations}, via a concret example we demonstrate how to utilize SSG representations to analysis symmetries and physical properties of materials. 

\section{Projective reps of anti-unitary groups}\label{anti-unitary representations}

In this section, we delve into the representation theory of anti-unitary groups, with a particular focus on projective co-representations. 
While SSGs may be either unitary or anti-unitary, the unitary case is considered a trivial condition of the anti-unitary scenario. 
Therefore, our attention is primarily directed towards understanding the anti-unitary condition.

An anti-unitary group can be represented as
\begin{equation}\label{T0}
    G = H + T_0 H,
\end{equation}
where $H$ denotes the maximal unitary subgroup of $G$, and $T_0$ is a fixed but arbitrarily chosen anti-unitary element.
For a projective representation $M(g)$ of an element $g \in G$, given that $g_1 g_2 = g_3$, the multiplication relationship can be expressed as
\begin{equation}
M(g_1)\kappa_{s(g_1)}M(g_2)\kappa_{s(g_2)}=\omega(g_1,g_2)M(g_3)  \kappa_{s(g_3)} ,
    \label{anti multiple}
\end{equation}
where $s(g)=1, \kappa_{s(g)}=I$ if $g$ is unitary and $s(g)=-1, \kappa_{s(g)}=\kappa$ (here and later $\kappa$ stands for complex conjugation) if $g$ is antiunitary. 
The U(1) factor $\omega(g_1, g_2)$ is a complex number with unit modulus, along with all instances of $\omega(g_1 \in G, g_2 \in G)$, constitutes the factor system of the group, 
adhering to the equation
\begin{equation}
    \omega(g_1, g_2) \omega(g_1g_2, g_3) = \omega(g_2,g_3)^{s(g_1)} \omega(g_1, g_2g_3).
    \label{factor anti}
\end{equation}

Two factor systems $\omega$ and $\omega'$ corresponding to two projective reps $M(g)$ and $M'(g)$ are considered equivalent if there exists a gauge transformation $\Omega(g) \in U(1)$, where $g \in G$, that applies to the rep $M$ such that
\begin{equation}
   M'(g)\kappa_{s(g)} = \Omega(g)M(g)\kappa_{s(g)}.
\end{equation}
In this case, the two factor systems are related by the following equation:
\begin{equation}
    \omega'(g_1, g_2) = \frac{\Omega(g_1)\Omega(g_2)^{s(g_1)}}{\Omega(g_1 g_2)}\omega(g_1, g_2).
\end{equation}

All reps that satisfy Equation~(\ref{factor anti}) qualify as permitted representations of a given group characterized by a specific factor system. 
However, our analysis will primarily focus on irreps, alongside how to reduce a reducible representation. 
The reducibility of a projective representation $M(g)$ of a finite anti-unitary group $G$ can be determined by the following criterion\cite{yang_hamiltonian_2021, kim1984}:
\begin{equation}
    \frac{1}{|H|} \sum_{h \in H} \frac{1}{2} \left[ \chi(h) \chi^*(h) + \text{Tr}[M(T_0 h) M^*(T_0 h)] \right] = 1,
\end{equation}
where $|H|$ denotes the order of the group $H$ and $\chi(g) = \text{Tr}[M(g)]$ denotes the trace of the representation matrix for an element $g\in G$.
If we define the `norm' of an irrep of an anti-unitary group as\cite{Shaw1974_1,Shaw1974_2}:
\begin{equation}
    R = \frac{1}{|H|} \sum_{h \in H}|\chi(h)|^2.
\end{equation}
Then the real, complex, and quaternionic classes or irreps\cite{Shaw1974_2,yang_hamiltonian_2021} have the norm $R=1, 2, 4$, respectively. 

When a reducible rep $M^r(g)$ is transformed into the reduced form as a direct sum of irreps, the multiplicity that an irrep $M^i(g)$ appears is equal to
\begin{equation}
    a_i = \frac{1}{R|H|}\sum_{h}\chi^i (h)^* \chi^r(h)  ={\sum_{h}\chi^i (h)^* \chi^r(h)\over \sum_{h }|\chi^i(h)|^2},
    \label{multiplicity}
\end{equation}
where $\chi^i(h)$ and $\chi^r(h)$ represent the characters of the irreducible and reducible reps, respectively. 
The process illustrates the complexity in reducing reps of anti-unitary group compared to those of unitary groups, given the variability of the norm among the irreducible components. It is worth mentioning that for groups without anti-unitary elements, these equations reduce to the standard forms found in textbooks, where the norm of each irrep consistently equals to 1. 

Next, we demonstrate that for any anti-unitary group, all of its projective irreps can be derived from its projective regular rep. 
The 
projective regular rep 
with a factor system $\omega$ is defined by acting the group elements on the group space itself, namely, 
\begin{equation}
 M_{ab}(g) = \langle g_a |\hat{g}|g_b\rangle = \omega(g,g_b)\delta(gg_b, g_a),
\end{equation}
with $g$, $g_a$ and $g_b$ being specific group elements. This expression is readily verifiable through Equation~(\ref{anti multiple}).
Similar to the regular reps of unitary groups, the degree of the anti-unitary projective regular rep is also equal to the order of the group.  
Each row and column contains only one non-zero element, but unlike linear regular reps where the element is always equal to 1, 
here the nonzero entry equals to the factor $\omega(g,g_b)\in U(1)$ with $g,g_b\in G$. Since the regular rep is constructed based on the group multiplication law, it implies that the identity element is always represented by the identity matrix, while non-identity elements necessarily have rep matrices with zeros on their diagonals, meaning that their characters must be 0.
By applying Equation~(\ref{multiplicity}), we calculate the multiplicity that an irreps $M^q$ appeared in the regular rep 
using the dimensionality $l_q$ and 
the norm $R_q$, as follows:
\begin{equation}
    a_q = \frac{2 |H| l_q}{R_q |H|}=\frac{2l_q}{R_q} \in \mathbb Z.
\end{equation}
Obviously, $a_q \geq 1$, which shows that all projective irreps can be obtained by reducing the projective regular rep, similar to the situations in the regular linear rep of unitary groups. 

It is known that the number of nonequivalent linear irreps equals the number of classes of the group \( G \). For an arbitrary anti-unitary group \( G = H + T_0H \), the classes of \( G \) consist solely of unitary elements. Two unitary elements, \( h_1 \) and \( h_2 \), belong to the same class if 
\[
\exists h \in H, \ \left( h h_1 h^{-1} = h_2 \text{ or } (h T_0) h_1^{-1} (h T_0)^{-1} = h_2 \right).
\]
In projective representation theory, a class can be further distinguished by a property called \(\omega\)-regularity, which holds if, for each element \( x \) in the class, 
\[
\omega(c,x) = \omega(x,c),\  \text{ for all } c\in G  \text{ such that } cx = xc.
\]
Actually, the number of nonequivalent projective irreps equals to the number of $\omega$-regular classes\cite{Yang2017rep, Karpilovsky_Chap3, Dijkgraaf-Witten}. 


We have established a procedure to construct all irreps of an anti-unitary group with a fixed factor system $\omega$. 
This involves constructing its regular rep, reducing it, and then identifying non-equivalent irreps. 
Reducing a rep into its irreducible components is a technical task, with various methods available, including the use of Complete Sets of Commuting Operators (CSCO)\cite{CSCO_method, Yang2017rep, ren2023enumeration} and the Hamiltonian method\cite{yang_hamiltonian_2021}. 
In this paper, we employ the Hamiltonian method to reduce the regular rep. The mechanism is summarized in the following theorem,  
\begin{theorem}
For an $d$-dimensional unitary rep $M(G)$ of the group $G$, 
construct the following matrix
\[
\mathscr H = \sum_{g \in G} \big[M(g)\kappa_{s(g)}\big] (X + X^\dag) \big[M(g)\kappa_{s(g)}\big]^{-1},
\]
in the rep space with \( X \) an $d$-dimensional {\bf random} matrix. 
Then the Hermitian matrix $\mathscr H$ commutes with $M(G)$, and the orthonormal bases of each eigen-space of $\mathscr H$ 
carry an unitary irrep that is contained in $M(G)$.
\end{theorem}

This method also applies for unitary groups. For detailed algorithm of this approach, see the Appendix~\ref{sm:hamiltonian method}.

\section{Little groups and Little co-groups of SSGs}\label{little group}
We introduce some fundamental types of groups which are essential for understanding the reps of SSGs: the full SSG, the little group of the SSG, and the little co-group of the SSG. These terms can be confusing, so we'll clarify them one by one.

First we review the definition of SSGs (i.e. the full SSG).
In this paper, we adopt the notation for SSGs as presented in Reference~\cite{jiang2023enumeration}, which is expressed as
\begin{equation}
     \mathcal G^{(S)}=\bigcup_{i=1}^{n} \{U_i || R_i|\bm{\tau}_i\} \bm{T},
     \label{SSG_definition}
\end{equation}
where $U_i\in \text{O}(3)$,  $R_i$ is a lattice point operation, $\bm{\tau}_i$ is a fractional translation, and $n$ is the number of cosets according to the translation group $\bm T$. 
When $\det(U_i)=-1$, it is assumed that it contains TRS $\mathcal{T}$ and is anti-unitary, 
i.e., $U_i\sim \det(U_i)U_i\cdot \mathcal{T}^{1-\det(U_i)\over2}$. 
Under this assumption, 
we do not distinguish $U_i\in \text{SO}(3)\times Z_2^T$ and $U_i\in \text{O}(3)$, and the full operation of $g\in \mathcal G^{(S)}$ reads
\begin{equation}
g=\left\{
\begin{aligned}
    &\{U || R_i|\bm{\tau}_i\}, ~\text{if }\det(U)=+1\\  
    &\{-U || R_i|\bm{\tau}_i\}\mathcal{T}, ~\text{if }\det(U)=-1.
\end{aligned}
\right.
\end{equation}
This notion is convenient to describe the symmetry of the magnetic structures,
which are often characterized by the vector field $\pmb M(\bm{r})$, showing a magnetic moment $M$ at position $\bm{r}$.
A SSG operation acts on the magnetic moment field $\pmb M(\bm{r})$ as
\begin{equation}
    \{U_i || R_i|\bm{\tau}_i\}\pmb M(\bm{r}) = U_i\pmb M (\{R_i|\bm{\tau}_i\}^{-1}\bm{r}).
\end{equation}
In the context of magnetic structures, a SSG operation is defined as a symmetry operation if it preserves the configuration of the magnetic moment field $\pmb M(\bm{r})$ invariant. 
This method is exactly how we determine the symmetry of magnetic structures, 
whether within MSG or SSG, by identifying operations that maintain the magnetic moment field.

The group $\mathcal{S}_0$ formed by pure-spin symmetry operations $\{U || E|\bm{0}\} \in \mathcal G^{(S)}$, 
which is called the `spin-only group', is an invariant subgroup of $\mathcal G^{(S)}$.
And we can define the quotient group $\mathcal G^{(S)}/\mathcal{S}_0$.
Based on the dimensional characteristics of magnetic structures, 
spin-only groups are systematically classified into four distinct classes:
\begin{enumerate}
    \item \textbf{Non-magnetic}: $\mathcal{S}_0=\text{O}(3)$, and the quotient group is just its SG;
    \item \textbf{Collinear magnetic order}:$\bm{M}(\bm{r})=(0, 0, \text{M}_z(\bm{r}))$ are set to along $z$-direction without loss of generality. 
    In this case, the spin-only group 
    \begin{equation}
        \mathcal{S}_0=\{\mathscr C^z_{\infty}|| E|\bm{0}\} + \{M_x \mathscr C^z_{\infty}|| \mathcal T|\bm{0}\} \cong \text{O}(2),
        \label{Eq:collinear_S0}
    \end{equation}
    and the spin part of the quotient group $\mathcal G^{(S)}/\mathcal{S}_0$ can only be 1 and -1.
    \item  \textbf{Coplanar magnetic order}:Also without loss of generality, 
    $\bm{M}(\bm{r})=(\text{M}_x(\bm{r}), \text{M}_y(\bm{r}), 0)$ are set to lie on the $z=0$ plane.
    In this case, 
    \begin{equation}
        \mathcal{S}_0=\mathbb{Z}_2^{M_z}=\{E, \{M_z || \mathcal T|\bm{0}\}\}
        \label{Eq:coplanar_S0}
    \end{equation}
    where $M_z$ denotes the mirror along $z$-axis.
    \item \textbf{Non-coplanar magnetic order}:$\mathcal{S}_0=E$ and the quotient group is just equal to the SSG.
\end{enumerate}

We then turn to the little group of the SSG at a specific $\bm{k}$ point. 
The little group of SSG $G(k)$ is defined as
\begin{equation}
    G(\bm{k}) = \{g\in \mathcal G^{(S)} | g\bm{k} - \bm{k} = \bm{K}\},
\end{equation}
where $\bm{K}$ is a reciprocal lattice vector.
It is important to note that the action of an SSG operation on the $\bm{k}$ vector encompasses two parts: the lattice space operation $(R^{-1})^T \bm{k}$ and the anti-unitary operation determined by $\det(U) \bm{k}$. This leads to the fact that the little group of an SSG does not always contain an intact spin-only group, particularly if the $\bm{k}$ point is not a time-reversal invariant momentum (TRIM) point. In cases of coplanar and collinear SSGs, the spin-only group includes an anti-unitary component that transforms $\bm{k}$ to $-\bm{k}$. 
In this context, both the $\bm{k}$ point and the reciprocal lattice vector $\bm{K}$ are defined with respect to the Brillouin zone established by the pure translation operations $\{E || E | a_1, a_2, a_3\}$, rather than the spin Brillouin zone (SBZ) referred to in some research\cite{xiao2023spin}.

Both the full SSG $\mathcal G^{(S)}$ and the little group $G(k)$ are infinite groups, making it challenging to construct their reps. Therefore, we introduce the concept of the little co-groups of SSGs $L(k)$. For MSGs and SGs, the little co-group is the point group of the little group, which is isomorphic to a Magnetic Point Group (MPG) or a Point Group (PG). However, the SSG has some differences.

We define the little co-group of the SSG as the little group of the SSG modulo the pure translation group, i.e., $L(\bm{k}) = G(\bm{k})/\bm{T}$. 
More specifically, $L(\bm{k}) = $
\begin{equation}\label{littleco}
\Big\{g \in \bigcup_{i=1}^{n} \{U_i || R_i\} \mid \{U_i || R_i|\bm{\tau}_i\}\in \mathcal G^{(S)}, g\bm{k} - \bm{k} = \bm{K}\Big\}.
\end{equation}

The $L(\bm{k})$ is not necessarily isomorphic to what is commonly referred to as a Spin Point Group (SPG)\cite{Litvin1974,schiff2023spin}. 
This discrepancy arises because the spin translation component of the SSG, $\{U||E|\bm{\tau}\}$, corresponds to the point group $\{U||E\}$, which represents a spin-only operation. 
For SPGs, any non-trivial spin-only group falls into the same classification as for SSGs. 
Consequently, the point group of the little co-group of the SSG may include a spin-only group composed of the spin part's point group, which differs from a SPG. 
The multiplicity relation of the $L(\bm{k})$ is
\begin{equation}
   \{U_i ||  R_i\}\{U_j ||  R_j\}= \{U_i U_j ||  R_i R_j\}.
   \label{mul relationship-lcg}
\end{equation}

\section{Representation of SSG}\label{rep of SSG}

The translation group is a normal subgroup of a SSG, and the number of cosets is generally finite (except for collinear SSGs). The irreps of the translation group are all 1-dimensional $D^{\bm{k}}(\bm{r}) = e^{-i\bm{k} \cdot \bm{r}}$ and are characterized by momentum points in the Brillouin zone. Generally, a momentum point $\pmb k$ is transformed into new momenta under the action of the cosets, and the set of these momentum points form a $k$-star ($k^*$). An irrep of the SSG preserves the $k^*$ and contains the information of diagonal part and off-diagonal part for the $k$ points in $k^*$. The most important information is actually contained in the diagonal part, especially the representation of the little co-group $L(\bm{k})$. The irrep of the $L(\bm{k})$ determines the degeneracy of energy band for single particles. Therefore, in the following we focus on the irreps of $L(\bm{k})$.


If $L(\bm{k})$ is anti-unitary, then the derivation of its irreps need the group structure (including the halving unitary subgroup and an antiunitary representative $T_0$), the multiplication relations, and the factor system determined by the spin and sub-lattice degrees of freedom. 

\begin{figure*}[htbp]
    \centering
    \includegraphics[width=1\textwidth]{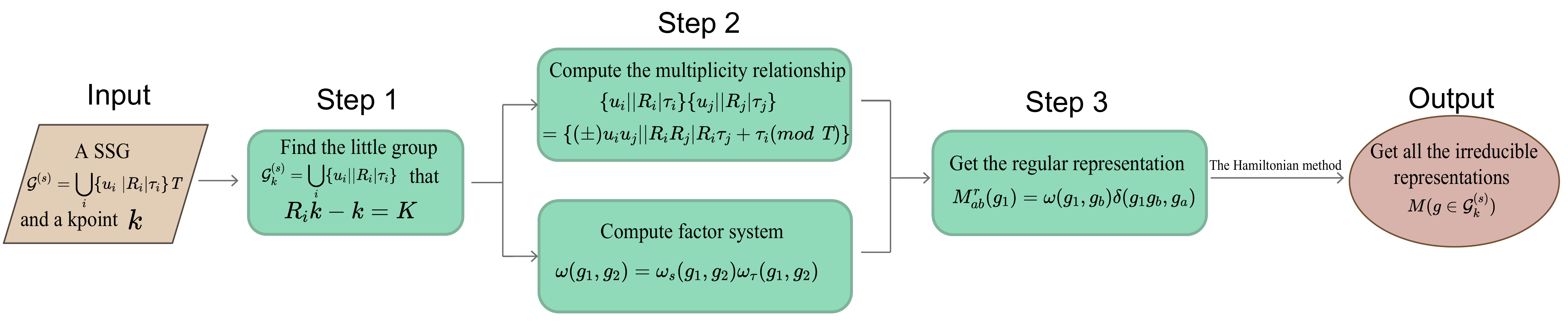}
    \caption{\label{flowchart} Flowchart of the algorithm for getting irreps of SSG. 
    }
\end{figure*}




Here we outline in detail how to obtain the necessary information for $L(\bm{k})$ and construct all their representations. 
We begin with the spin-only group $\mathcal{S}_0$, whose specific form 
has been detailed in the section~\ref{little group}. Notice that a SSG always has the group structure  \cite{jiang2023enumeration}
\[
\mathcal{G}^{(S)} = \mathcal{S}_0 \times \left(\mathcal{G}^{(S)}/\mathcal{S}_0\right).
\] 
The cases involving coplanar and non-coplanar situations are relatively straightforward where $\mathcal{S}_0$ is a finite group, but 
the collinear cases need special treatment.
Collinear spin-only group is an infinite group $\mathcal{S}_0 = \{\mathscr C^z_{\infty}|| E|\bm{0}\} + \{M_x \mathscr C^z_{\infty}|| \mathcal T|\bm{0}\}$. 
Due to the conservation of $S_z$, a conventional approach is to quotient the infinite group $\mathscr C^z_{\infty}$ away by considering the spin-up (or spin-down) components only.
The quotient group for the spin-up (spin-down) component effectively becomes type-II MSG, with $\{M_x|| \mathcal T|\bm{0}\}$ as an effective time-reversal operation $T_0= \{C_{2x}\mathcal T|| \mathcal T|\bm{0}\}$, 
with $T_0^2=1$\cite{ren2023enumeration,xiao2023spin}. 
While this strategy aptly describes the symmetry of the spin-up (spin-down) component independently, it fails to account for up-down degeneracy. 
To remedy this issue, instead of completely ignoring $\mathscr C^z_{\infty}$ we replace it with its subgroup $\mathscr C^z_{2}\subset \mathscr C^z_{\infty}$. Namely, we replace the infinite spin-only group $\mathcal{S}_0$ by its finite subgroup $\mathcal{S}'_0 = \{\mathscr C^z_{2}|| E|\bm{0}\} + \{M_x \mathscr C^z_{2}|| \mathcal T|\bm{0}\}\subset \mathcal{S}_0$. Accordingly, we obtain a subgroup of the original SSG, namely, 
\[
\mathcal{G}^{'(S)} = \mathcal{S'}_0 \times \left(\mathcal{G}^{(S)}/\mathcal{S}_0\right)\subset \mathcal{G}^{(S)}.
\]
Once the irreps of the subgroup $\mathcal{G}^{'(S)}$ are obtained, one can extend them to the irreps of the original SSG $\mathcal{G}^{(S)}$ without changing the dimensionality. The efficacy of this finite subgroup in characterizing the infinite collinear SSG will be elucidated in the Appendix~\ref{sm:collinear}.



The little co-group can be obtained from Equation ~(\ref{littleco}). Notice that the spin-only group generally does not belong to 
$G(\bm{k})$ unless $\bm{k}$ is a TRIM point. This necessitates extracting the elements of the little group directly from the full SSG ($\mathcal{G}^{(S)}$) rather than solely from the quotient group ($\mathcal{G}^{(S)}/S_0$). 
Furthermore, when considering collinear conditions, the continuous rotational symmetry group $\{\mathscr C^z_{\infty }||E\}$ invariably remains an invariant subgroup of $G(\bm{k})$. This presence underscores the importance of our strategy to replace it by its finite subgroup.

Then we analyze the factor system of SSGs.
The well-discussed concept of group isomorphism primarily hinges on the first two aspects: the unitary-anti-unitary relationship and multiplication relations. 
Consequently, all SSGs can be associated with a corresponding isomorphic MSG, as documented in some literature\cite{xiao2023spin}. 
Nevertheless, despite their isomorphism, the distinction in factor systems endows them with markedly different representations. 
This underpins the broader spectrum of representations available to SSGs compared to MSGs. 
In the following discussion, we will concentrate on the factor system of $L(\bm{k})$. 
This system is derived from two primary sources: the $\text{SU}(2)$ factor $\omega_s$ and the non-symmorphic translation factor $\omega_\tau$.

First we show the origin of the $\text{SU}(2)$ factor $\omega_s$. The SSG operation acts on the spinor wave function as:
\begin{equation}
    \begin{aligned}
        &\{U_i ||  R_i|\bm{\tau}_i\} \begin{pmatrix}
       \psi_\uparrow (\bm r)\\\psi_\downarrow (\bm r)
     \end{pmatrix} =u_i\begin{pmatrix}
       \psi_\uparrow ( \{R_i|\bm{\tau}_i\}^{-1} \bm r)\\\psi_\downarrow (\{R_i|\bm{\tau}_i\}^{-1}\bm r)
     \end{pmatrix},
     \end{aligned}
\end{equation}
where $u_i$ is the $\text{SU}(2)$ matrix corresponding to $U_i$(and contains an anti-unitary operation when $\text{det}U_i < 0$).
However, due to the $1:2$ homomorphism between the $SO(3)$ group and the $\text{SU}(2)$ group, $\pm u_i$ both corresponds to the same $U_i$ . 
To address the complexities arising from this relationship, a common strategy is to use double groups.
This approach treats the -1 factor of the $\text{SU}(2)$ component acts as a new group element $\bar{E}$, leading to $^d\{U_i || R_i|\bm{\tau}_i\} = \bar{E} \{U_i || R_i|\bm{\tau}_i\}$. 
Doubling the group elements ensures that all elements still follow linear multiplication relations.

This concept 
has been applied in the representation of MSGs, where the choice between single and double group representation hinges on whether the element doubling induced by $\text{SU}(2)$ is incorporated. 
For magnetic systems having SSG symmetry, 
double groups (according to the spin operations) should be chosen in the analysis of the electron energy bands, while single groups should be adopted to study the magnon bands. 

However, introducing double groups doubles the number of group elements, significantly increasing the computational effort required for representation calculations.
Consequently, we choose to assimilate the additional -1 factor introduced by $\text{SU}(2)$ into the inter-element factors. 
For instance, consider three elements $g_1$, $g_2$, and $g_3$, for which $u(g_1)u(g_2)= -u(g_3)$. 
In double group terminology, their multiplicity relation is expressed as $g_1g_2=\bar{E}g_3$. Here we have $g_1g_2=-1\times g_3$, with -1 serving as the factor $\omega_s$ between $g_1$ and $g_2$.


Moreover, even for single non-symmorphic SSGs (also true for non-symmorphic SGs and MSGs), the linear irreps of the little group are characterized by projective irreps of the little co-group when the translation groups are quotient out. The factor system $\omega_\tau$ of the projective of the little co-group comes from the sub-lattice degrees of freedom, namely, the non-symmorphic translations. 
For unitary group elements $g_1,g_2,$ the factor $\omega_\tau(g_1,g_2)$ arising from non-symmorphic translation is commonly chosen as:
\begin{equation}
    \omega_\tau(g_1,g_2) = e^{-i \bm{K_{1}} \cdot \bm{\tau}_2},
\end{equation}
where $\bm{K_{1}} = g_1^{-1} \bm{k} - \bm{k}$. This facilitates the connection between projective representation $M(R)$ of the little co-group and linear representation $D(\{R|\bm{\tau}\})$ of the little group, namely, 
\begin{equation}
    M(R) = \exp(i\bm{k}\cdot \bm{\tau}) D(\{R|\bm{\tau}\}). 
    \label{linear2projective}
\end{equation}
It can be easily proved that this factor $\omega_\tau$ adheres to the combination relationship required for projective representations.

Given that most SSGs are anti-unitary, we extend this formulation to encompass the anti-unitary condition. 
Let the matrix relationship between linear representation and projective representation still satisfies equation~(\ref{linear2projective}),
we can deduce the non-symmorphic translation factor $\omega_\tau(g_1,g_2)$ of an anti-unitary group\cite{yang2021factor}:
\begin{equation}
    \omega_\tau(g_1,g_2) = e^{-i \bm{K_{1}} \cdot \bm{\bm{\tau}_2}},
    \label{tau factor}
\end{equation}
where $\bm{K_{1}} =s(g_1)( g_1^{-1} k - k)$ and $s(g_1) = 1$ when $g_1$ is unitary and $s(g_1)=-1$ when anti-unitary. 
This form of factor also satisfies the combination relationship of anti-unitary group with the specific proof presented in the Appendix~\ref{prove of factor tau}.

For quasiparticle carrying half-integer spin (like the electrons), 
one should consider both the spin degrees of freedom and non-symmorphic translations, the factor system $\omega(g_1,g_2)$ of the projective reps of the little co-group at $\bm k$ is a product of $\omega_s$ and $\omega_\tau$,
\begin{eqnarray}
\omega(g_1,g_2) = \omega_s(g_1,g_2)\omega_\tau(g_1,g_2).
\end{eqnarray}

As an example, consider an $L(\bm{k})$ of SSG 1.4.1.3 at the $\bm{k}$ point $(0,0,\pi)$, 
whose non-symmorphic translation and factor system from $\text{SU}(2)$ (for electrons) are both non-trivial. 
Four elements make up this SSG: two unitary elements, $\{E||E|000\}$ and $\{C_{2z}||E|011\}$, and two anti-unitary elements, $\{M_{x}||\mathcal T|010\}$ and $\{M_{y}||\mathcal T|001\}$. 
When transforming the basis into the unit cell of SSG, four elements being $\{E||E|000\}$ and $\{C_{2z}||E|0\frac{1}{2}\frac{1}{2}\}$, $\{M_{y}||\mathcal T|00\frac{1}{2}\}$ and $\{M_{x}||\mathcal T|0\frac{1}{2}0\}$. 
Using the spin part for convenience, we label them as $E$, $C_{2z}$, $M_{y}$, and $M_{x}$, respectively, and we then examine the factor between $M_{y}$ and $M_{x}$. 
The $\omega_s$ of the multiplicity connection $M_y M_x = C_{2z}$ is first determined. The $\text{SU}(2)$ matrices that correspond to them are:
\begin{equation}
    u(M_y) = -i\sigma_y \kappa u(C_{2y}) = \begin{pmatrix}
        -1&0\\0&-1
    \end{pmatrix}  \kappa,
\end{equation}
\begin{equation}
    u(M_x) = -i\sigma_y \kappa u(C_{2x}) = \begin{pmatrix}
        -i&0\\0&i
    \end{pmatrix} \kappa,
\end{equation}
\begin{equation}
    u(C_{2z})=\begin{pmatrix}
-i&0\\0&i
    \end{pmatrix}.
\end{equation}
Notice the anti-unitary operation $\kappa$ that $\kappa i = -i \kappa$ here, we get the $\omega_s$ between them as:
\begin{equation}
    \omega_s(M_y,M_x)=1\;,\omega_s(M_x,M_y) = -1.
\end{equation}

\begin{table}[t]
    \begin{tabular}{c|c|c|c|c}
    \hline\hline
    elements & \;$E$\;       & $C_{2z}$ & $M_{y}$& $M_{x}$  \\\hline
    $E$      & 1 ,1 ,1       & 1 ,1 ,1 & 1 ,1 ,1 &1 ,1 ,1             \\\hline
    $C_{2z}$   & 1, 1, 1     & -1, 1, -1 & -1, 1, -1&1, 1, 1          \\\hline
    $M_{y}$   & 1 ,1 ,1      & 1, -1, -1 & 1, -1, -1&1, 1, 1      \\\hline
    $M_{x}$      &1 ,1 ,1    & -1, -1, 1 & -1, -1, 1&1 ,1 ,1     \\
    \hline\hline
    \end{tabular}
    \caption{\label{tab:factor of 1.4.1.3} The factor system of SSG 1.4.1.3 at k point $(0,0,\pi)$. Three mumbers in the table represent the factors $\omega_s$, $\omega_\tau$ and $\omega = \omega_s \omega_\tau$ (for half-integer spin) associated with the left and upper elements, respectively.}
    \end{table}

Next we turn to the non-symmorphic translation factor $\omega_\tau$. This can be deduced by just using the equation~(\ref{tau factor}) that 
\begin{equation}
    \omega_\tau(M_y, M_x) = e^{-i \bm{K}_{M_y} \cdot \bm{\tau}_{M_x}}.
\end{equation}
In the unit cell of SSG, $\bm{\tau}_{M_x} = (0, 0.5, 0)$ and $\bm{K}_{M_y} = (0, 0, 2\pi)$, so we get $\omega_\tau(M_y, M_x) = 1$ is trivial.
But since $\bm{\tau}_{M_y} = (0, 0, 0.5)$ and $\bm{K}_{M_x} = (0, 0, 2\pi)$, $\omega_\tau(M_x, M_y) = -1$.
Lastly , we can multiple them to get the factor system $\omega = \omega_s \omega_\tau$ for fermionic quasiparticles. The whole factor system of the little group is shown in Tab.~\ref{tab:factor of 1.4.1.3}.

\begin{table}[htbp]
    \begin{tabular}{c|c|c|c|c}
    \hline\hline
    elements & \;$E$\;       & $C_{2z}$ & $M_{y}$& $M_{x}$  \\\hline
\multirow{2}{*}{single group }       &$\begin{pmatrix}  1&0 \\  0&1 \end{pmatrix}$    & $\begin{pmatrix}  1&0 \\  0&-1 \end{pmatrix}$  & $\begin{pmatrix}  0&1 \\  -1&0 \end{pmatrix}\kappa$   &$\begin{pmatrix}  0&1 \\  1&0 \end{pmatrix}\kappa$     \\\hline
\multirow{2}{*}{double group }  &$\begin{pmatrix}  1&0 \\  0&1 \end{pmatrix}$    & $\begin{pmatrix}  -i&0 \\  0&i \end{pmatrix}$  & $\begin{pmatrix}  0&1 \\  -1&0 \end{pmatrix}\kappa$   &$\begin{pmatrix}  0&i \\  i&0 \end{pmatrix}\kappa$     \\
    \hline\hline
    \end{tabular}
    \caption{ \label{tab:irreps of 1.4.1.3} Irreps at (0, 0, $\pi$) for the single group of 1.4.1.3 (for integer spin having factor system $\omega_\tau$) and the double group of 1.4.1.3 (for half-integer spin having factor system $\omega=\omega_s\omega_\tau$). }
    \end{table}

After obtainning the factor system, we can easily construct the projective regular representation of the $L(\bm{k})$. 
And in this example, regular representation will be reduced into two equivalent representations,
which means this little group has only one irrep with dimension of two. 
The resultant irreps for the single group of 1.4.1.3 and the double group of 1.4.1.3 are listed in Tab.~\ref{tab:irreps of 1.4.1.3}.

Until now, we have delineated the procedure to obtain the reps of SSGs, illustrated via a flowchart in Fig.~\ref{flowchart}. 
It is important to note that the reps achieved through this procedure are projective, rather than linear. The projective irreps of an SSG correspond to linear irreps of the corresponding double SSG. 

\section{Applications of SSG}\label{ultilities of SSG representations}

\subsection{Genaral Discussion}

The SSGs and their irreps can be employed to systematically investigate the electronic strctures and physical properties of magnetic materials exhibiting weak SOC. To this end, we have done the following four steps to facilitate the application of SSGs:

(a) \textbf{Determination of SSG Operations}: Ascertain the complete set of SSG operations of a given magnetic structure, 
which includes both quotient and spin-only groups. The identification of the SSG group number is facilitated by referencing the established SSG database.

(b) \textbf{Extraction of irreps at $k$-points}: Enumerate all distinct irreps of the SSG's little co-groups at various $k$-points, 
focusing on high-symmetry points and lines. 

(c) \textbf{Characterization of Band Structures}: If band data is available from first-principle calculations, it is possible to classify each band utilizing irreps. 
Previously, tools such as IRVSP\cite{GAO2021irvsp} were employed to characterize bands in SG and MSG contexts for non-magnetic and magnetic materials, respectively. 
The key here is to compute the traces of electronic states\cite{Vergniory2019} and extending this methodology to SSGs facilitates the analysis of band structures in magnetic materials exhibiting weak SOC, including antiferromagnetic variants. 
Some band degeneracies, not explicable by MSG, are elucidated by SSGs. Through representation theory, it is possible to ascertain which specific operations protect these degeneracies.

(d) \textbf{
Construction of $k\cdot p$ Theory}: With the matrix representations of SSG in hand, their application to $k\cdot p$ theory to develop effective Hamiltonians is a natural progression. 
There are already mature programs that can derive $k\cdot p$ basis vectors starting from representation matrices, applicable to SGs and MSGs, these methods are equally valid for SSGs. 
However, due to SSG's enhanced symmetry, leading to an increased number of representation matrices and a rise in representation dimensions, the computational cost of these traditional methods could become significantly high. 
We also introduce a novel approach that offers certain advantages for handling high-dimensional representations in the Appendix~\ref{sm:kp}. 

To assist researchers in applying this methodology to the study of other materials, we have developed and organized the entire process—including the identification of SSG, obtaining the specific little group representations of the SSG, and deriving independent $k\cdot p$ Hamiltonians from the little group representations at relevant k-points—into code. This code has been uploaded to GitHub(\url{https://github.com/zine-phy/MOM2SSG} and \url{https://github.com/zine-phy/SSGReps}) for use.

\subsection{A Concerete Example: Mn$_3$Sn}

Then we use the well-discussed coplanar anti-ferromagnetic Weyl semi-metal with weak SOC, \ch{Mn3Sn}\cite{Li2017_Mn3Sn,Markou_Mn3Sn,Liu2017_Mn3Sn,Chen2021_Mn3Sn,Hiho2018_Mn3Sn,Yang_2017_Mn3Sn,Taylor2020_Mn3Sn}, which has a hexagonal lattice with Mn atoms forming two Kagome sublayers stacked along the c-axis, as an example to analyze its symmetry.
The magnetic moments of \ch{Mn} atoms are oriented in the $(1,1,0)$, $(-1,0,0)$ and $(0,-1,0)$ directions, respectively, written under $\bm{a}_1, \bm{a}_2, \bm{a}_3$ axes, 
The algorithm to find all the SSG permitted symmetry operations and the results of \ch{Mn3Sn} are shown in the Appendix~\ref{sm:algorithm of band representations}.
Using the label in the SSG dataset, its SSG number reads 194.1.6.1.P, where the P letter means its coplanar magnetic structure.
The corresponding MSG is 63.463 (BNS number), which exhibits much lower symmetry compared to its SSG.

\begin{figure}[htbp]
    \centering
    \includegraphics[width=0.48\textwidth]{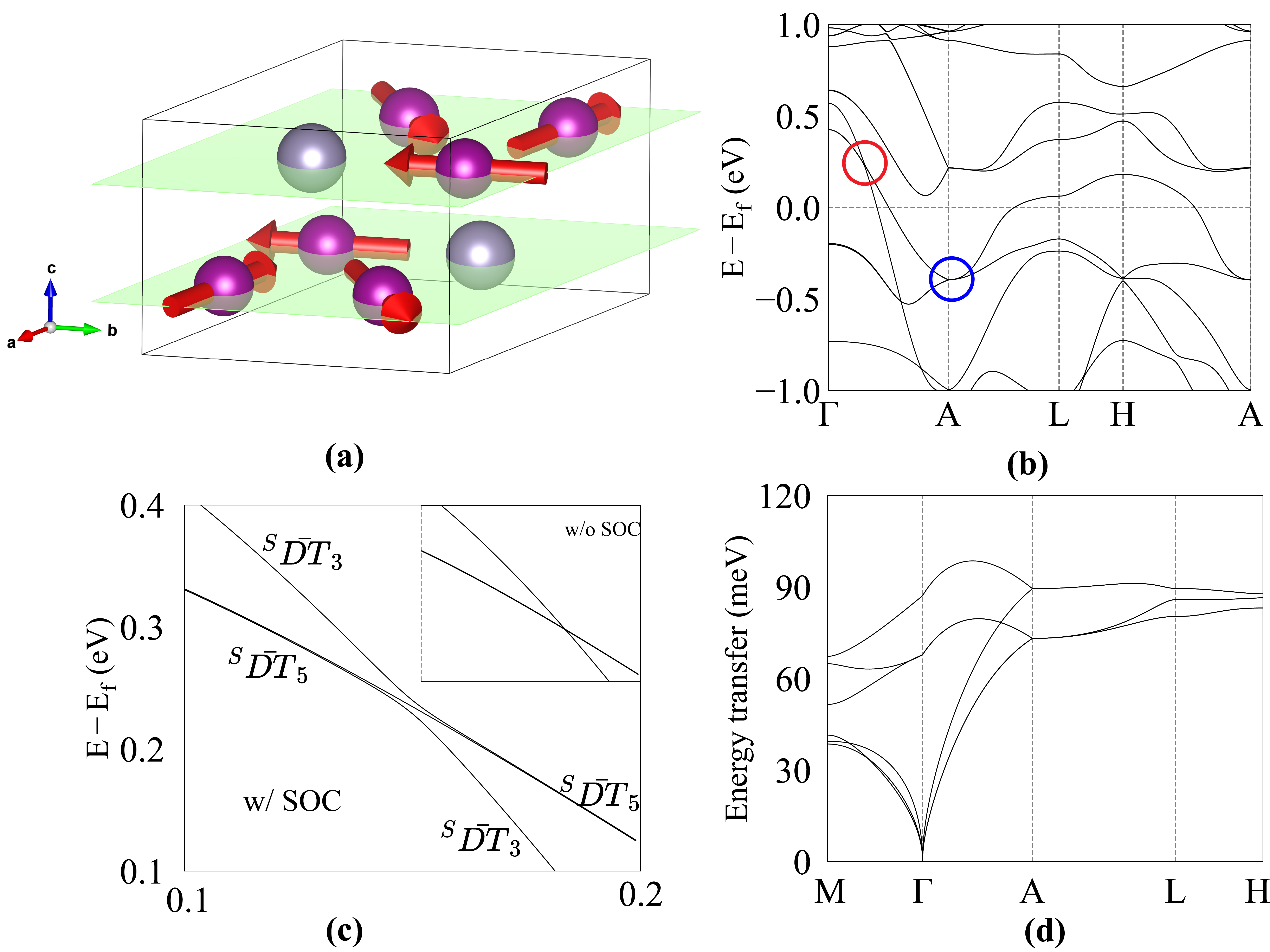}
    \caption{\label{Fig:band_representations} 
    (a) \ch{Mn3Sn} has a hexagonal lattice where Mn atoms form two Kagome layers and develop a coplanar magnetic order.
    (b) Band structure of \ch{Mn3Sn} near the fermi level along the k point path $\Gamma -A-L-H-A$ with SOC.
    (c) The enlarged band structure of \ch{Mn3Sn} near the Fermi level along the $\Gamma-A$ path with SOC. The inset in the upper right corner shows the band structure without SOC for comparison. related band representations of SSG are labeled.
    (d) Spin wave dispersion along the $M-\Gamma-A-L-H$ path.}
\end{figure}

\subsubsection{Band structure and degeneracies}

As illustrated in Fig.~\ref{Fig:band_representations}, we sketch its electronic band structure on the k path $\Gamma -A-L-H-A$ and compute the irreps of $L(\bm{k})$ on its high symmetry lines and high symmetry points.
The character tables are presented in the Appendix~\ref{sm:character tables}, and we label each representation as $^s \bar{X}_n$, where the number $n$ counts the representations starting at 1, the $S$ letter indicates the representation of SSG, and $X$ is the label of the k point.
We will demonstrate the band degeneracy protected by the SSG in the absence of SOC.
For materials with weak SOC, the degeneracy will be preserved if the band rep of the MSG retains the same dimensionality as that of the SSG. However, if the band rep of the MSG has a lower dimensionality, the degeneracy will be slightly lifted.

Without SOC, the generators of $L(\bm{k})$ along the $\Gamma-A$ line contain $\{2_{010}||M_{010}\}$, $\{3^-_{001}||3^+_{001}\}$, $\{E||2_{001}|z/2\}$ and the anti-unitary operation $\{M_z||P\mathcal T\}$.
As seen in Fig.~\ref{Fig:band_representations}(c),  a singly degenerate band (represented as $^s \bar{DT}_3$) intersects with a doubly degenerate band (represented as  $^s \bar{DT}_5$) to form a triple degeneracy point. This band crossing is precisely identifiable by the representations of the SSG along the path.
The linear dispersion near the triple degeneracy point can be captured by the  $k \cdot p$  Hamiltonian, which includes three independent parameters $a,b,c$ at the first order in $\bm{k}$, as shown in Appendix~\ref{sm:kp matrices}. Two of the parameters, $a, b$ are associated with $\delta k_z$ and are given by: 
\begin{equation}
    H^{(1)}_k = \begin{pmatrix}
  a&  0&0 \\
  0&b  &0  \\
  0& 0 & b 
\end{pmatrix} \delta k_z.
\end{equation}
The existence of 3-fold degeneracy, resulting from the crossing between the two inequivalent irreps, are accident from the viewpoint of representation theory but are stable from the viewpoint of topological bands. 

The $L(\bm{k})$ at the A point (located at the boundary of the Brillouin zone) includes all generators of the $L(\bm{k})$ along the $\Gamma - A$ line and an additional generator $\{ E||P \}$,
which contain 2d and 4d irreps. These irreps account for the observed 2d and 4d degeneracies at the $A$ point, as shown in Tab.~\ref{tab:ssg character A point}. 
We also employed the $k\cdot p$ method to analyze the dispersion around this Dirac point using the irreps of the SSG at the $A$ point. Our analysis revealed that there is only one independent parameter in the first-order $k\cdot p$ Hamiltonian, namely
\begin{equation}
    H^{(1)}_k = a(\sigma_z \otimes \sigma_0) \delta k_z,
\end{equation}
where $\sigma_z$ is a Pauli matrices and $\sigma_0$ is the $2\times 2$ identity matrix. This Hamiltonian accurately captures the linear dispersion along the $A-\Gamma$ path near the $A$ point, with $\sigma_0$ ensuring the 2d degeneracy away from the A point.

For the $A-L$ path, the analysis requires considering the $\delta k_x$ component.  The lowest order dispersion is quadratic, and the second order $k\cdot p$ Hamiltonian contains two independent parameters associating with $(\delta k_x)^2$,
\begin{equation}
\begin{aligned}
    &H^{(2)}_k(\delta k_y=\delta k_z=0) \\
    =&\left[ b 
    \begin{pmatrix}
    \sigma_0  & 0\\
     0 & \sigma_0
    \end{pmatrix} + c
    \begin{pmatrix}
    \sqrt{3}\sigma_1-\sigma_2  & 0\\
     0 & \sqrt{3}\sigma_1+\sigma_2
     \end{pmatrix}\right](\delta k_x)^2,
 \end{aligned}
\end{equation}
which agrees well with the quadratic dispersion observed along the $A-L$ path near the $A$ point. The full $k \cdot p$ Hamiltonian derived from the 4d SSG irrep is shown in Appendix~\ref{sm:kp matrices}. Similar behavior is observed for other Dirac points at the $A$ point, such as the one approximately 0.3 eV above the Fermi level. 

\subsubsection{The Effect of weak but finite SOC}
Now we analysis the band structure with nonzero SOC, which can be conducted by restricting the SSG to its subgroup MSG symmetry. 
Generally, when the symmetry operations that protect degeneracies are broken upon the introduction of SOC, the degeneracies will be lift with a gap whose the size is dependent on the strength of SOC.

When SOC is turned on, the little co-group along the $\Gamma-A$ path is a MPG $m'm2'$ whose generators include $M_{010}$ and $M_z \cdot \mathcal{T}$. This MPG only possesses two one-dimensional representations, $\bar{B}_3$ and $\bar{B}_4$. 
Notably, in the presence of SOC, away from the crossing point on the $\Gamma$-$A$ line the two energy levels according to $^s \bar{D}T_5$ remains almost degenerate, with a gap of only around 1 meV, affirming the validity of using SSG as a tool to approximately analyze the symmetries of \ch{Mn3Sn}.

Differences between the SOC and non-SOC scenarios emerge near the band crossing; the three-fold degeneracy point no longer appears, but rather the line of two-fold degeneracy below bifurcates and remixes with the band above to form a new line of two-fold degeneracy. This unusual band behavior warrants a combined analysis using the representations of both MSG and SSG. As a subgroup of SSG, MSG allows for the restriction of SSG representations onto it. The SSG representation  $^s \bar{DT}_3$ when restricted to MSG corresponds to $\bar{B}_3$, while $^s \bar{DT}_5$ is reducible and splits into two inequivalent irreps, $\bar{B}_3$ and $\bar{B}_4$. Near the band crossing, the two-fold degenerate line breaks into two bands represented as $\bar{B}_3$ and $\bar{B}_4$, which then recombine away from the crossing point to reform a nearly two-fold degenerate line protected by SSG. The gap between the two bands with the same rep $\bar{B}_3$ is approximately 12 meV, caused by the avoided band crossing between bands with identical reps.

With SOC, the little co-group at the A point is a MPG $m'm'm$, with generators $P$, $M_{010}$ and $M_z \cdot \mathcal{T}$.
Since this MPG lacks 4d irreps, the 4d degeneracy splits in general.
Along the $\Gamma-A$ line, 
this MPG only supports 1d irreps as we have shown above, 
and at the $A$ point, it supports 2d irreps.  
Consequently, the nodal lines and Dirac point develop a small gap of approximately 1 meV. 

Nonetheless, the nodal line along the $A-L$ path remains intact when turning on SOC, 
as the irreps continues to be 2d when restricted to MSG (the little co-group on the $A-L$ path reads MPG $2'2'2$). Furthermore, the MSG-permitted operation $T_0=\{M_z||C_{2z}\mathcal T|{{\bf \hat z}\over 2}\}$ protect at least 2d Kramers-type degeneracy across the entire $k_z=\pi$ plane, as it is an anti-unitary operation satisfying $\omega(T_0,T_0)=\omega_s(T_0,T_0)\omega_\tau(T_0,T_0)=-1$.\\

\subsubsection{Spin wave dispersion and degeneracies}
SSGs can also be applied to describe the symmetries of magnon band structures\cite{ssgmagnon2022}. Since magnons are bosons, the construction of their representations differs slightly. Specifically, the $\text{SU}(2)$ factor, which is relevant for fermionic systems, is absent in SSGs describing bosonic systems. To construct the representations for magnons, we follow a similar procedure as outlined for fermionic systems, with the key constraint that $\omega_s = 1$. In our online database, users can obtain the irreducible representations for bosonic systems by selecting `single group' under `Group Type'.

As an example, we analyze the magnon band structure of Mn$_3$Sn. 
We construct a Heisenberg model and calculate the magnon dispersion using SpinW\cite{Tothspinw}, a package for calculating magnetic excitations based on linear spin wave theory. 
Note that the focus here is on demonstrating the effectiveness of SSG in describing the degeneracies within the magnon band structure. 
Consequently, the $J_i$ parameters in the Heisenberg model are chosen to ensure the positive definiteness of the Hamiltonian, though they may not correspond to the actual values for the real material.

The resulting spin wave dispersion is displayed in Fig.~\ref{Fig:band_representations}(d). 
The structure of the little group is the same as what we previously discussed in the analysis of the electronic band structure. 
Along the $\Gamma - A$ path, its single group includes both 1D and 2D irreps, while at the $A$ point, 2D and 4D irreps are present. 
This accounts for the existence of nodal lines along the $\Gamma - A$ path and a 3D crossing point, which also appears in the electronic band structure. 
At the $A$ point, two nodal lines intersect to form a Dirac point, which carries a 4D irrep.

Another intriguing result is that similar to the electron band structure, there still exist an entire nodal plane at $k_z = \pi$ in the magnon band. 
At a general k-point within the $k_z = \pi$ plane, the little group includes the anti-unitary operation $T_0 = \{M_z || C_{2z}\mathcal T | {{\bf \hat z}\over 2}\}$. 
The non-symmorphic translation introduces a non-trivial factor $\omega_\tau(T_0, T_0) = -1$, making $T_0$ an anti-unitary operation with $T_0^2 = -1$, which results in Kramers degeneracy across the entire $k_z = \pi$ plane.

\section{conclusions and outlook}
In this work, we develop an algebra method to obtain all the irreps SSGs by focusing on the projective irreps of little co-group at any $k$ point which are generally anti-unitary spin point groups. We then develop a comprehensive methodology for applying SSGs to the study of actual magnetic materials. We believe that this foundational effort on SSGs will greatly enhance their applications in both theoretical and experimental endeavors in the future.

The method employed to obtain irreps starts from the fundamental information like the  group multiplication table and the factor system, making it 
applicable to across various types of groups.
Thus, this approach can also be utilized to calculate irreps of 
new groups beyond the present work, serving as a convenient mathematical tool.

It is demonstrated that the irreps of SSGs can be applied in band representation theory and $k\cdot p$ theory for quasiparticles around degenerate momentum points. 
The application of SSGs and their irreps can be further extended to identify the refined magnetic structures (via neutron scattering), and to develop selection rules in exploring the physical response of high-dimensional quasiparticles.
Therefore, the complete acquisition of SSG rep matrices is expected to significantly contribute to the advancement of these and related fields, providing a robust foundation for future research and development.

The method to obtain irreps of SSGs and some of their applications mentioned above have been coded into Python programs and are available in \href{https://github.com/zine-phy/MOM2SSG}{GitHub} or the \href{https://cmpdc.iphy.ac.cn/ssg/}{online Database}.

\acknowledgments{
This work was supported by the National Natural Science Foundation of China (Grants No. 11925408, No. 11921004, No. 12188101, No. 12374166, No. 12134020 and No. 12325404), the Ministry of Science and Technology of China (Grant No. 2022YFA1403800, No. 2023YFA1406500, and No. 2022YFA1405300), the Chinese Academy of Sciences (Grant No. XDB33000000) and the New Cornerstone Science Foundation through the XPLORER PRIZE.
}

\normalem
\bibliography{ref}

\clearpage
\newpage
\onecolumngrid

\begin{center}
\textbf{\center{\large{Supplementary Material}}}
\end{center}

\beginsupplement

\tableofcontents

\section{Reducing any given projective co-representations of anti-unitary groups}\label{sm:hamiltonian method}

Given an arbitrary projective co-representation \( M \) of an anti-unitary group \( G = H + T_0H \) with factor system $\omega_2$, where $H$ denotes the maximal unitary subgroup of $G$, and $T_0$ is a fixed but arbitrarily chosen anti-unitary element, a natural question arises: how can we find a unitary transformation $U$ that reduces \( M \) into a block-diagonal form, where each block represents an irreducible projective co-representation? For unitary groups, such $U$ had been explicitly constructed\cite{Serre1977}, but a simple construction to decompose any given projective representations of anti-unitary groups is lack\cite{Yang2017rep}.  
Here, we introduce a powerful way called the Hamiltonian approach\cite{yang_hamiltonian_2021} to make the construction explicitly. 

We start with a lemma. 
\begin{lemma}
If a hermitian matrix $Q$ commutes with $M(G)\kappa_{s(G)}$, then each eigen-space of $Q$ labels an invariant subspace of the representation $M(G)\kappa_{s(G)}$. 
\end{lemma}
The hermitian condition is necessary since the group $G$ is anti-unitary. The Hamiltonian approach focuses on the space \(\mathscr{Q}\) formed by all Hermitian centralizers of \( M(G)\kappa_{s(G)} \) whose coefficients take values in \(\mathbb{R}\). \(\mathscr{Q}\) can be seen as analogous to the space of \( G \)-homomorphisms for a unitary group. 
\begin{define}[Intertwining number for co-representation]
Let \( V \) be the carrier space for the representation \( M(g)\kappa_{s(G)} \). The intertwining number \( I(V, V) \) of \( V \) is defined as the dimension of \( \mathscr{Q} \), which can be calculated using the following formula:
\begin{align}
I(V, V)={1\over 2|H|} \sum_{h\in H}\left\{ \chi(h)\chi^*(h)  + \omega(T_0h, T_0h) \chi((T_0h)^2)\right\}.
\end{align}
\end{define}
The intertwining number of a space \( V \) equals 1 if and only if \( V \) is irreducible, and the basis for the one-dimensional space \( \mathscr{Q} \) is the identity matrix. 
When the intertwining number of a space \( V \) is greater than 1, a general matrix of the space \( \mathscr{Q} \) is not the identity matrix, and the decomposition of eigenspaces according to the different eigenvalues of that matrix leads to a reduction of the given representation. Hence, once the bases of the space \(\mathscr{Q}\) are known, step-by-step reductions can be performed to make intertwining number of all subspaces ultimately be 1. Furthermore, such a step-by-step procedure essentially makes full use of the \( I(V, V) \)-independent bases of the space \(\mathscr{Q}\): 
Firstly, the step-by-step reduction procedure yields a direct sum decomposition of the space \(\mathscr{Q}\):
\[
\mathscr{Q} = \mathscr{Q}_0 \oplus\mathscr{Q}_1 \oplus \mathscr{Q}_2 \oplus \cdots \oplus \mathscr{Q}_n,
\]
where \(\mathscr{Q}_0\) consists of matrices that are proportional to the identity matrix, and \(\mathscr{Q}_i, i=1,\dots,n\), represents the set of matrices associated with the \(i\)-th reduction step, where the dimension of \(\mathscr{Q}_i\) is greater than 1, and the \(i\)-th reduction commutes with the reductions from steps \(\{1, \dots, i-1\}\). Therefore, by definition, each \(\mathscr{Q}_i, i=1,\dots,n\) consists of all matrices that, when projected onto subspaces obtained from the \(i\)-th step, are proportional to the identity matrix. Secondly, for each step, a general matrix is chosen in \( \mathscr{Q}_i \) to decompose the eigenspaces according to the different eigenvalues of that matrix. Such a general matrix can be obtained by summing all independent bases of the space \( \mathscr{Q}_i \) with random coefficients in \(\mathbb{R}\). Thirdly, if we consider such randomly constructed matrices throughout the entire procedure, since every \( i \)-th step commutes with all \(\{1, \ldots, i-1\}\) steps, it is equivalent to saying that the whole reduction procedure is carried out by a single matrix formed by summing all original \( I(V, V) \)-independent bases of the space \(\mathscr{Q}\) with random coefficients in \(\mathbb{R}\).
Hence, we prove the following theorem:
\begin{theorem}
Let \( Q_1, Q_2, \ldots, Q_i \) denote the bases of the space \(\mathscr{Q}\) formed by all Hermitian centralizers of \( M(G)\kappa_{s(G)} \), and let \( r_1, \ldots, r_n \) be a series of random coefficients in \(\mathbb{R}\). Then the eigenspaces of the following matrix
\[
Q_r = \sum_{l=1}^{i} r_l Q_l
\]
provide projectors that project the representation space \( V \) into a direct sum of irreducible components.
\end{theorem}

The bases of the space \(\mathscr{Q}\) can be obtained using the Hamiltonian approach, similar to the general \( k \cdot p \) theory discussed in Sec\ref{sm:kp}. Here we mention an alternative method, which is equivalent to the above procedure and more effective in numerical calculations, to obtain the desired \( Q_r \).
\begin{corollary}
Let \( Q_0 \) be a random matrix with the same dimension as \( V \). Construct the following matrix:
\[
q_r = \sum_{g \in G} [M(g)\kappa_{s(g)}] Q_0 [M(g)\kappa_{s(g)}]^{-1}.
\]
Then the eigenspaces of the matrix
\[
Q_r = q_r + q_r^\dag
\]
provide projectors that project the representation space \( V \) into a direct sum of irreducible components.
\end{corollary}

\section{Changing spin-only group for representations of collinear SSGs}\label{sm:collinear}
For systems with collinear magnetic structures, when SOC is neglected, spin remains a good quantum number due to the $ SO(2)$ symmetry corresponding to rotations around the z-axis. A common strategy is to treat the spin-up and spin-down components separately, both yielding the symmetries $\{E||H\}$ and $\{M_x||H\}$, where $H$ represents pure lattice operations, and $M_x \equiv C_{2x} \bm{T}$ is anti-unitary. 
The advantage of this method is that we can directly use the single-valued representations($\text{SU}(2)$ factor $\omega_s$ always be 1) of the grep group of $H$ which is access on the Bilbao.
Here the choice of these single group representations stems from the $\text{SU}(2)$ factor 
\begin{equation}
    \omega_s(\{M_x||E\}, \{M_x||E\}) = -i\sigma_y \kappa e^{-i\frac{\pi}{2}\sigma_x}\times(-i\sigma_y \kappa e^{-i\frac{\pi}{2}\sigma_x}) = 1.
\end{equation} 
However, this method cannot account for the up-down degeneracy, which is crucial for distinguishing between antiferromagnetic and altermagnetic structures.

For ferromagnetic structures, the full symmetry of the SSG that they possess is given by:
\begin{equation}
    \{E||H\},\; \{M_x||H\},\;\{\mathscr C^z_{\infty}||H\}.
\end{equation}
The first two parts form a finite group, which corresponds to the grep group we previously mentioned. In fact, $\{\mathscr{C}^z_{\infty}||H\}$ does not affect the degeneracy since $[\{M_x||E\},\;\{C_{z\theta}||E\}] = 0$, where $\theta$ is an arbitrary angle.

The conditions for antiferromagnetic structures are significantly more complicated, and the full symmetry operations include:
\begin{equation}
    \{E||H\},\; \{M_x||H\},\;\{\mathscr C^z_{\infty}||H\},\;\{\bm{T}||A\}\;\{C_{2x}||A\},\;\{\mathscr C^z_{\infty}\bm{T}||A\},
\end{equation}
where $A$ contains real-space transformations which interchange atoms between opposite-spin sublattices.
Now we can no longer neglect $\mathscr C^z_{\infty}$ since  it introduces extra degeneracy. For 
\begin{equation}
    e^{-i\frac{\pi}{2}\sigma_x}e^{-i\frac{\theta}{2}\sigma_z} = e^{-i\frac{-\theta}{2}\sigma_z}e^{-i\frac{\pi}{2}\sigma_x},
\end{equation}
these terms do not commute with each other.
This degeneracy can also be captured using $\mathscr C^z_{2}$, avoiding the need for the infinite group $\mathscr C^z_{\infty}$.
From the perspective of group representation theory, the representation space for  $\mathscr C^z_{2}$ is also closed under the action of $\mathscr C^z_{\infty}$, making it sufficient to capture the full symmetry of the group. 
If one prefers, using either $\mathscr C^z_{3}$ or $\mathscr C^n_{z}$ would also work. We find that $\mathscr C^z_{2}$ is sufficient to capture the degeneracy while minimizing the group's order.

\section{The non-symmorphic translation factor for anti-unitary group}\label{prove of factor tau}
To verify the factor form is always 
\begin{equation}
    \omega_\tau(g_1,g_2) = e^{-i \bm{K_{1}} \cdot \bm{\bm{\tau}_2}},\; \text{where}\; \bm{K_{1}} =s(g_1)( g_1^{-1} k - k),
\end{equation}
there still two conditions need to verify, both are anti-unitary and the second operation is anti-unitary.
The verification is very similar so we will won't give them here.

We revisit the derivation of the non-symmorphic translation factor $\omega_\tau$ for SG. 
The relationship between linear representations $D(g)$ and projective representations $M(g)$ is defined as:
\begin{equation}
    M(\{R|\bm{\tau}\}) = \exp(ik\cdot \bm{\tau}) D(\{R|\bm{\tau}\}).
\end{equation}
From this relation, alongside the multiplication relationship of linear representations:
\begin{equation}
    D(\{R_1|\bm{\tau}_1\})D(\{R_2|\bm{\tau}_2\})=D(\{R_1 R_2|R_1\bm{\tau}_2 + \bm{\tau}_1\}),
\end{equation}
we derive the factor between SG elements as:
\begin{equation}
    \begin{aligned}
    M(R_1|\bm{\tau}_1)M(R_2|\bm{\tau}_2)&=\exp[ik\cdot (\bm{\tau}_1+\bm{\tau}_2)]D(R_1|\bm{\tau}_1)D(R_2|\bm{\tau}_2)\\
    &= \exp[ik\cdot (\bm{\tau}_1+\bm{\tau}_2)]D(R|\bm{\tau})\exp[-ik\cdot(R_1\bm{\tau}_2+\bm{\tau}_1-\bm{\tau})]\\
    &= \exp[ik\cdot (\bm{\tau}_1+\bm{\tau}_2)]M(R|\bm{\tau})\exp[-ik\cdot \bm{\tau}]\exp[-ik\cdot(R_1\bm{\tau}_2+\bm{\tau}_1-\bm{\tau})]\\
    & = \exp[ik\cdot (\bm{\tau}_2 - R_1 \bm{\tau}_2)]M(R|\bm{\tau})\\
    & = \exp[-i(R_1^{-1}k - k)\cdot \bm{\tau}_2]M(R|\bm{\tau})\\
    & \equiv  \exp[-iK_r\cdot \bm{\tau}_2]M(R|\bm{\tau})
    \end{aligned}
\end{equation}
where $R = R_1 R_2$ and $\bm{\tau} = R_1\bm{\tau}_2 + \bm{\tau}_1 \mod 1$. 
This formulation is similarly applied in anti-unitary groups, but when deriving $\omega_\tau$, special attention must be given to the action of anti-unitary operations on the factor. 
Here, we omit the $\text{SU}(2)$ component as our focus is on the translational part. 
Assuming $\{R_1|\bm{\tau}_1\}$ is anti-unitary and $\{R_2|\bm{\tau}_2\}$ is unitary, the factor $\omega_\tau$ between them is expressed as:
\begin{equation}
    \begin{aligned}
    M(R_1|\bm{\tau}_1)\kappa M(R_2|\bm{\tau}_2)&=\exp[ik\cdot (\bm{\tau}_1-\bm{\tau}_2)]D(R_1|\bm{\tau}_1)\kappa D(R_2|\bm{\tau}_2)\\
    &= \exp[ik\cdot (\bm{\tau}_1-\bm{\tau}_2)]\exp[-ik\cdot(R_1\bm{\tau}_2+\bm{\tau}_1-\bm{\tau})] D(R|\bm{\tau})\kappa \\
    &= \exp[ik\cdot (\bm{\tau}_1-\bm{\tau}_2)]M(R|\bm{\tau})\exp[-ik\cdot \bm{\tau}]\exp[-ik\cdot(R_1\bm{\tau}_2+\bm{\tau}_1-\bm{\tau})]\kappa\\
    & = \exp[-ik\cdot (\bm{\tau}_2 + R_1 \bm{\tau}_2)]M(R|\bm{\tau})\kappa\\
    & = \exp[-i(R_1^{-1}k + k)\cdot \bm{\tau}_2]M(R|\bm{\tau})\kappa\\
    & \equiv  \exp[-iK_r\cdot \bm{\tau}_2]M(R|\bm{\tau})\kappa.
    \end{aligned}
\end{equation}
To affirm that the factor form consistently remains
\begin{equation}
    \omega_\tau(g_1,g_2) = e^{-i \bm{K_{1}} \cdot \bm{\bm{\tau}_2}},\; \text{where}\; \bm{K_{1}} =s(g_1)( g_1^{-1} k - k),
\end{equation}
two additional conditions (the second operation being anti-unitary and both operations being anti-unitary) should be verified. 
The verification process, being similar to the above one, will not be detailed here.

\section{Algorithm for identifying SSG and band representation}\label{sm:algorithm of band representations}

The algorithm for identifying the SSG associated with a magnetic material is structured into two main phases: 
(i) the enumeration of all SSG operations applicable to the magnetic structure, and (ii) the determination of the corresponding SSG number within our database. 
Here we will focus on the first step here since getting all the symmetry operations is enough for getting representations and the labeling of SSG is not included in this paper.

To enumerate all SSG operations pertinent to the material, our algorithm first segregates all feasible real-space operations and spin rotations. 
It then iterates over the potential combinations of these operations to identify those that preserve the symmetry of the magnetic material. 
This process effectively filters out the operations that leave the magnetic structure invariant, thereby constituting the SSG specific to the material. 
To delineate the algorithm in detail:
\begin{enumerate}
\item \textbf{Real-space Operations Identification}: Initially, we identify all possible real-space operations by examining the atomic positions and species, 
whilst disregarding the magnetic moments. This task is efficiently executed using the \textit{spglib} package.
\item \textbf{Spin Rotations Enumeration}: Subsequently, we determine all possible spin rotations by identifying the point group (PG) that the magnetic moments' orientations in spin space constitute. 
Analogous to points defining a 'molecular' structure in spin space, the aggregation of magnetic moments forms a comparable structure whose PG is ascertainable with tools like \textit{pymatgen}. 
We mention a few technical details in this step:
\begin{itemize}
\item An additional point at the origin of spin space is introduced to anchor the reference point, thereby avoiding redundant operations.
\item Initial identification of collinear or coplanar magnetic orders simplifies the subsequent steps. 
Specifically, for collinear orders, the spin-only-free group is limited to the identity and a possible mirror operation. 
For coplanar orders, a manual adjustment aligns the magnetic moment plane to approximate non-coplanar configurations, effectively isolating the spin-only group formed by the mirror operations.
\end{itemize}
\end{enumerate}

Here we also use the \ch{Mn3Sn} to show the effectiveness of this algorithm.

(a) When neglect the magnetic moments of the \ch{Mn3Sn}, we identify its SG as $P6_3/mmc$. 
All the lattice operation $\{R_i|\bm{\tau}_i\}$ of the SSG operations $\{u_i||R_i|\bm{\tau}_i\}$ must be within the SG $P6_3/mmc$.

(b) \ch{Mn3Sn} has six Mn atoms in a single cell, and these Mn atoms have three different magnetic moment orientations, $(-3, 0, 0)$, $(1.5, 2.598, 0)$ and $(1.5, -2.598, 0)$, which all within the xy plane. 
Since Mn3Sn is a coplanar magnetic material, we add a z-component to all these magnetic moments to isolate the spin-only group.
Thus we get four points after add a point at the origin:
\begin{equation}
    (0,0,0), (-3, 0, m_z), (1.5, 2.598, m_z), (1.5, -2.598, m_z),
\end{equation}
where $m_z$ can be an arbitrary non-zero number. We identify the PG of the 'molecular' formed by this four points as $3m$.

(c) Combine the operations of SG $P6_3/mmc$ and PG $3m$ to form the lattice operations and spin operations of a SSG operations, i.e. $\{3m||E\}\otimes\{E||P6_3/mmc\}$.
This group's subgroup has to include the material's SSG. Therefore, all we have to do is evaluate each component to see if it retains its magnetic structural identity.
Regarding the \ch{Mn3Sn}, it is seen that every SG element retains its magnetic structure identity when combined with a PG element, indicating that this magnetic structure has the highest degree of symmetry in the SSG field.

It's important to note that this approach can also be used to locate MSG operations and the only additional requirement is that the spin and lattice operations' rotational components be locked.
This approach will be also helpful in evaluating the anti-ferromagnetic and altermagnetic materials, which have received a lot of attention lately.

\section{Charactor tables of 194.1.6.1.P at high symmetry points and lines}\label{sm:character tables}
\begin{table}[htbp]
    \begin{tabular}{c|c|c|c|c|c}
    \hline\hline
    SSG operations & $\{E||E\}$ & $\{E||\mathcal{P}\}$ & $\{2_{010}||M_{010}\}$ & $\{3^{-}_{001}||3^+_{001}\}$ & $\{E||2_{001}|\frac{z}{2}\}$ \\\hline
    $^S \bar{\Gamma}_1\rule{0pt}{2.6ex}$            & 1 & -1 & -i & -1 & -1             \\\hline
    $^S \bar{\Gamma}_2\rule{0pt}{2.6ex}$            & 1 & -1 &  i & -1 & -1           \\\hline
    $^S \bar{\Gamma}_3\rule{0pt}{2.6ex}$            & 1 &  1 &  i & -1 & -1         \\\hline
    $^S \bar{\Gamma}_4\rule{0pt}{2.6ex}$            & 1 &  1 & -i & -1 & -1           \\\hline
    $^S \bar{\Gamma}_5\rule{0pt}{2.6ex}$            & 1 & -1 & -i & -1 &  1           \\\hline
    $^S \bar{\Gamma}_6\rule{0pt}{2.6ex}$            & 1 & -1 &  i & -1 &  1           \\\hline
    $^S \bar{\Gamma}_7\rule{0pt}{2.6ex}$            & 1 &  1 &  i & -1 &  1           \\\hline
    $^S \bar{\Gamma}_8\rule{0pt}{2.6ex}$           & 1 &  1 & -i & -1 &  1           \\\hline
    $^S \bar{\Gamma}_9\rule{0pt}{2.6ex}$            & 2 & -2 & 0 & 1 & -2           \\\hline
    $^S \bar{\Gamma}_{10}\rule{0pt}{2.6ex}$           & 2 &  2 & 0 & 1 & -2           \\\hline
    $^S \bar{\Gamma}_{11}\rule{0pt}{2.6ex}$           & 2 & -2 & 0 & 1 &  2           \\\hline
    $^S \bar{\Gamma}_{12}\rule{0pt}{2.6ex}$           & 2 &  2 & 0 & 1 &  2  \\
    \hline\hline
    \end{tabular}
    \caption{\label{tab:ssg character gamma point} Character table of SSG 194.1.6.1.P. at $\Gamma$ point.}
\end{table}

\begin{table}[htbp]
    \begin{tabular}{c|c|c|c|c}
    \hline\hline
    SSG operations & $\{E||E\}$ &  $\{2_{010}||M_{010}\}$ & $\{3^{-}_{001}||3^+_{001}\}$ & $\{E||2_{001}|\frac{z}{2}\}$ \\\hline
    $^S \bar{DT}_1\rule{0pt}{2.6ex}$            & 1 & i  & -1 &$ -e^{-i\pi w}$              \\\hline
    $^S \bar{DT}_2\rule{0pt}{2.6ex}$            & 1 & -i & -1 &$ -e^{-i\pi w}$          \\\hline
    $^S \bar{DT}_3\rule{0pt}{2.6ex}$            & 1 & i  & -1 &$  e^{-i\pi w}$         \\\hline
    $^S \bar{DT}_4\rule{0pt}{2.6ex}$            & 1 & -i  & -1 &$ e^{-i\pi w}$        \\\hline
    $^S \bar{DT}_5\rule{0pt}{2.6ex}$            & 2 & 0 & 1 & $ -2e^{-i\pi w}$           \\\hline
    $^S \bar{DT}_6\rule{0pt}{2.6ex}$            & 2 & 0 & 1 & $ 2e^{-i\pi w}$           \\
    \hline\hline
    \end{tabular}
    \caption{\label{tab:ssg character DT point} Character table of SSG 194.1.6.1.P. on $(0,0,w)$ point, on the $\Gamma-A$ point.}
\end{table}


\begin{table}[htbp]
    \begin{tabular}{c|c|c|c|c|c}
    \hline\hline
    IRREPs & $\;E\;$ & $\{E||\mathcal{P}\}$ & $\{2_{010}||M_{010}\}$ & $\{3^{-}_{001}||3^+_{001}\}$ & $\{E||2_{001}|\frac{z}{2}\}$ \\\hline
    $^S \bar{A}_{1}\rule{0pt}{2.6ex}$            & 2 & 0 & 2i & -2 & 0             \\\hline 
    $^S \bar{A}_{2}\rule{0pt}{2.6ex}$           & 2 & 0 & -2i & -2 &  0           \\\hline
    $^S \bar{A}_{3}\rule{0pt}{2.6ex}$           & 4 &  0 & 0 & 2 &  0  \\
    \hline\hline
    \end{tabular}
    \caption{\label{tab:ssg character A point} Character table of SSG 194.1.6.1.P. at $A$ point.}
\end{table}

\begin{table}[htbp]
    \begin{tabular}{c|c|c|c}
    \hline\hline
    IRREPs & $\;E\;$  & $\{2_{010}||M_{010}\}$  & $\{E||2_{001}|\frac{z}{2}\}$ \\\hline
    $^S \bar{R}_{1}\rule{0pt}{2.6ex}$            & 2  & -2i & 0             \\\hline 
    $^S \bar{R}_{2}\rule{0pt}{2.6ex}$           & 2 & 2i  &  0           \\
    \hline\hline
    \end{tabular}
    \caption{\label{tab:ssg character R point} Character table of SSG 194.1.6.1.P. at $(w,0,0.5)$ point, on the $A-L$.}
\end{table}

\begin{table}[htbp]
    \begin{tabular}{c|c|c|c|c|c}
    \hline\hline
    IRREPs & $\;E\;$ & $\{E||\mathcal{P}\}$ & $\{2_{010}||M_{010}\}$  & $\{E||M_{001}|\frac{z}{2}\}$ \\\hline
    $^S \bar{L}_{1}\rule{0pt}{2.6ex}$            & 2 & 0 & -2i & 0          \\\hline 
    $^S \bar{L}_{2}\rule{0pt}{2.6ex}$           & 2 & 0 & 2i & 0         \\
    \hline\hline
    \end{tabular}
    \caption{\label{tab:ssg character L point} Character table of SSG 194.1.6.1.P. at $L$ point.}
\end{table}

\begin{table}[htbp]
    \begin{tabular}{c|c|c|c|c|c}
    \hline\hline
    IRREPs & $\;E\;$  & $\{2_{010}||2_{010}\}$ & $\{3^{-}_{001}||3^+_{001}\}$  & $\{E||M_{001}|\frac{z}{2}\}$ \\\hline
    $^S \bar{H}_{1}\rule{0pt}{2.6ex}$            & 2 & 0 & -2 & 0          \\\hline 
    $^S \bar{H}_{2}\rule{0pt}{2.6ex}$            & 2 & 0 & 1 & 0          \\\hline 
    $^S \bar{H}_{3}\rule{0pt}{2.6ex}$           & 2 & 0 & 0 & 0         \\
    \hline\hline
    \end{tabular}
    \caption{\label{tab:ssg character H point} Character table of SSG 194.1.6.1.P. at $H$ point.}
\end{table}

\section{Algorithm for $k\cdot p$ theory}\label{sm:kp}

At momentum \( k \), given an arbitrary projective representation \( M \) carried by basis $\Phi_k$ of the little co-group \(G\), we can derive the corresponding \( k \cdot p \) theory using a highly efficient method called the Hamiltonian approach\cite{yang_hamiltonian_2021}.

In general, the effective low-energy BdG Hamiltonian reads
\[
H_{\rm eff}=\sum_{\delta k} \Phi_{k+\delta k}^\dag\Lambda ({\delta k})\Phi_{ k+\delta k}=\sum_{\delta k}H_{ k+\delta k},
\]
where $\Lambda(\delta k)$ is a Hermitian matrix $\Lambda^\dag(\delta k)=\Lambda(\delta k)$. When summing over all the momentum variations, the total Hamiltonian should preserve the $G$ symmetry, namely
$\hat g\left(\sum_{\delta k}H_{ k+\delta k}\right)\hat g^{-1}=\left(\sum_{k}H_{k+\delta k}\right),$ for all $ g\in G$.  Assuming $\hat g\Phi^\dag (k+\delta k)\hat g^{-1} =  \Phi^\dag ( k+g\delta k) M(g)\kappa_{s(g)} $, then generally for any $g\in G$ one has
\beq\label{gsrh0}
M(g) \kappa_{s(g)} \Lambda \left( \delta k\right) \kappa_{s(g)} M^{\dagger}(g)=\Lambda(g \delta k),
\eeq

Starting from (\ref{gsrh0}), we derive the formula to judge the dispersion relation in momentum space.  For instance, we consider linear dispersion around $k$, namely
\beq
\Lambda(\delta k)=\sum_{m=1}^d \delta k_m\Lambda^m+O(\delta k^2).
\eeq
Here $\delta k$ is a dual vector under the action of group $G$, namely
\(
\hat g\delta k_m=\sum_n M_{mn}^{(\bar {\rm v})}(h)\delta k_n,
\)
where $(\bar {\rm v})$ is the dual Rep of the vector Rep $(\rm v)$ of the group $G$. 

{\it (I) $G$ is unitary.} If $G$ is a unitary group, then for any $g\in G$ one has
\begin{align}\label{varvZUni}
M(g) \Lambda^n  M(g)^\dagger  = \sum_{m} {M}_{mn}^{(\bar {\rm v})}(g)\Lambda^m
\end{align}
Thus, the existence of linear dispersion is determined by whether the product Rep $M(G)\otimes M^*(G)$ contains the linear Rep $M^{(\bar {\rm v})}(G)$ or not. It can be judged from the quantity
\begin{align}\label{Sec:kp20}
I_{(\rm \bar v)}={1\over |G|}\sum_{g\in G} |\chi(g)|^2\cdot\chi^{(\rm v)}(g) 
\end{align}
where $\chi^{({\rm v})}(g)\equiv\chi^{({\rm v})}(g) =\operatorname{Tr}[ {M}^{( \bar{\rm v})}(g)]^*$. If $I_{({\rm \bar v})} =0$, then the dispersion must be of order higher than 1. If $I_{({\rm \bar v})} \neq 0$, then the dispersion is linear, and one can always find  Hermitian matrices $\Lambda^m$ satisfying the equation (\ref{varvZUni}) noticing that the Rep $({\rm \bar v})$ is real.

{\it (II) $G$ is anti-unitary.}  The matrices $\Lambda^m$ also carry dual vector rep of the group $G$. 

For any $g\in G$, one has
\beq\label{varvZ}
M(g) \kappa_{s(g)}\Lambda^n \kappa_{s(g)} M(g)^\dagger  = \sum_{m} \Lambda^m {M}_{mn}^{(\bar {\rm v})}(g),
\eeq
where ${M}^{(\bar {\rm v})}(g) \kappa_{s(g)}$ is a real Rep of $G$ hence the operator $\kappa_{s(g)}$ can be hidden. Hence the existence of linear dispersion also depends on whether the product Rep $M(G)\otimes M^*(G)$ contains the Rep $M^{(\bar {\rm v})}(G)$ or not, under the condition that the CG coefficients can be reshaped into hermitian matrices.

Select an anti-unitary operation $T_0\in G$ and denote $G=H+T_0H$. Introduce the following matrices
$$
\tilde\Lambda^m=\Lambda^m\cdot M^{\rm T}(T_0).
$$ 
According to the Hamiltonian approach\cite{yang_hamiltonian_2021}, the symmetry condition (\ref{varvZ}) with $g=T_0$ and the hermitian condition $(\Lambda^m)^\dag =\Lambda^m$ are combined into a single symmetry condition of $\tilde\Lambda^m$ (called $\eta_0$-symmetry condition): 
\begin{align}\label{eta0-condition}
\big(\tilde\Lambda^m\big)^{\rm T}=\eta_0 D(T_0^2)\sum_n D^{(\bar {\rm v})}_{nm}(T_0)\tilde\Lambda^n.
\end{align}
with $\eta_0=\omega(T_0,T_0)$.

Therefore, when restricted to the $\eta_0$-symmetric subspace, one only needs to consider the unitary symmetries of $\tilde\Lambda^m$ as discussed in case (I). This is a relatively simpler since the unitary elements are represented in the $\mathbb{C}$ field. On the other hand, $\Lambda^m$ carry real representations of $G$ so one need to treat them in the $\mathbb{R}$ field which is more complicated. This subtlety makes the Hamiltonian approach a method with a high efficiency.

Similar to Eq.(\ref{Sec:kp20}), the existence of linear dispersion can be judged by the following quantity
\begin{align}\label{DispersionCriterion}
I_{({\rm \bar v})} =\frac{1}{|G|} \sum_{u\in H }\Big[|\chi(u)|^{2} \chi^{({\rm v})}(u) + \omega\left(u T_0, uT_0\right) \chi\big(\left(u T_0\right)^{2}\big) \chi^{({\rm v})}\left(u T_0\right) \Big],
\end{align} 
namely if $I_{({\rm \bar v})} \neq 0$, the dispersion is linear, otherwise the dispersion is of order higher than 1. Notice that the character $\chi^{({\rm v})}\left(u T_0\right)$ in (\ref{DispersionCriterion}) is well-defined although $uT_0$ is anti-unitary, because the rep $({\rm v})$ is a real rep such that the allowed bases transformations can only be real orthogonal matrices which keep $\chi^{({\rm v})}\left(u T_0\right)$ invariant.

Now we discuss how to obtain the matrices $\Lambda^m$ or $\tilde \Lambda^m$ in the case $I_{({\rm \bar v})} \neq 0$.  Besides the $\eta_0$-symmetry condition (\ref{eta0-condition}), the symmetry constraints for unitary elements $u\in G$ reads
\begin{align}
M(u)\tilde\Lambda^mR^{\rm T}(u)&=\sum_n M^{(\bar {\rm v})}_{nm}(u)\tilde\Lambda^n,\label{F-condition}
\end{align}
with $R(u) = M(T_0) M^*(u) M^\dag(T_0).$
If we consider the set of matrices $\tilde\Lambda^m$ as a single column vector $\tilde{\pmb\Lambda}$ with
$(\tilde{\pmb\Lambda})_{(n-1) \times d^2+(\alpha-1)\times d+\beta}=(\tilde\Lambda^n)_{\alpha\beta},$
then the above constrains (\ref{F-condition}) indicates that $\tilde{\pmb\Lambda}$ carries identity rep of any unitary element $u\in G$, namely 
\beq\label{WGm}
W(u)\tilde{\pmb\Lambda}=\tilde{\pmb\Lambda},
\eeq
with 
$
W(u)=M^{({\rm v})}(u)\otimes M(u)\otimes R(u).
$
Furthermore, $W(T_0)\kappa \tilde{\pmb\Lambda}=\tilde{\pmb\Lambda}$ also holds where $T_0$ is represented as $W(T_0)\kappa = M^{({\rm v})}(T_0)\otimes M(T_0)\otimes M(T_0)\kappa$.

When projected to the $\eta_0$-symmetric space, the rep $W(u)$ becomes
$W_{\eta_0}(u)\equiv {\rm{Proj}_{\eta_0}}W(u) $ with the matrix entries given by
\Beq
\!\!\! &&[W_{\eta_0}(u)]_{m\gamma\rho,n\alpha\beta}={1\over2}\Big[ M_{mn}^{({\rm v})}(u) M_{\gamma\alpha}(u) R_{\rho\beta}(u) + 
\\&&\ \ \ \ \eta_0 \sum_\lambda M_{mn}^{({\rm v})}(uT_0) M_{\gamma\beta}(u) R_{\rho \lambda}(u)M_{\lambda\alpha}\left((T_0)^2\right) \Big]. 
\Eeq
The above matrix $W_{\eta_0}(u)$ is not a rep for the group formed by all unitary elements $u\in G$, but it does contain the identity rep for that group, which shows that the dispersion contains linear term. Then the $\Lambda^m$ can be obtained via the following procedure: \\
(1) Obtain the common eigenvectors of $W_{\eta_0}(u)$ with eigenvalue 1 for all unitary elements $u\in G$. These eigenvectors span a Hilbert subspace $\mathcal L^{(I)}$. \\
(2) Project $W(T_0)K$ into $\mathcal L^{(I)}$, and perform bases transformation such that element $T_0$ is represented as $IK$ in $\mathcal L^{(I)}$. Then the new bases are all of the allowed $\tilde{\pmb \Lambda}$. \\
(3) Reshape each of the new bases into three matrices $\tilde \Lambda^x, \tilde \Lambda^y, \tilde \Lambda^z$. \\
(4) Finally one has $\Lambda^m=\tilde \Lambda^m [M(T_0)^{\rm T}]^{-1}$.\\

Similarly, one can obtain higher-order \( k \cdot p \) terms via the above procedure. We remark that the procedure can be applied to the 0-th order \( k \cdot p \) terms, {\it i.e.}, to obtain the bases of the space formed by all Hermitian centralizers of \( M(G)\kappa_{s(G)} \) as discussed in Sec.~\ref{sm:hamiltonian method}.

\section{Optimization of SSG Representation Matrices}\label{sm:optimization}

In the main sections of our paper, we discuss utilizing the Hamiltonian method to derive irreps. However, a minor issue persists, though it may not be crucial for most SSG applications: the representation matrices are not as elegant as desired. These matrices often contain arbitrary elements derived from the Hamiltonian method, lacking the concise diagonal or anti-diagonal structure that would make them more aesthetically pleasing.

In this supplementary section, we introduce a methodology for refining the presentation of representation matrices. Specifically, we outline three key steps in the optimization process: defining the standard form for projective co-representations, diagonalizing the representation matrices, and resolving any phase ambiguities.

\subsection{Standard form for irreducible co-representations}

First, it is essential to revisit the three types of anti-unitary projective representations that are critical for the optimization process.
An anti-unitary group is denoted as $G=H+T_0H$, where $H$ represents the maximum unitary subgroup and $T_0$ is an arbitrarily chosen anti-unitary element.
Consider $U(g)\kappa_{s(g)}$ as a projective irrep of $G$ with factor system $\omega$. This projective representation can be restricted to its maximum unitary subgroup $H$, resulting in the restricted rep $U(h),\forall h\in H$, which may be either reducible or irreducible\cite{Shaw1974_2,AZY2024_prb}. 
If irreducible, this representation is classified as type I.
If reducible, expressed as $U(h)=E^*(h) \oplus D(h)$, where $E^*$ and $D$ are both irreps of $H$, and if they are linearly equivalent, it is classified as type II.
If not, it is classified as type III.
Those familiar with induced co-representations from unitary representations may recognize these classifications, as types I, II, and III correspond to norms 1, 4, and 2, respectively. Therefore, without needing to examine the irreps of the unitary subgroup, it is straightforward to determine which type—I, II, or III—a given irreducible co-representation belongs to.

Upon clarifying the three types of co-representations, we introduce the concept of a standard form (often referred to as a `canonical form' in the literature\cite{Shaw1974_2}).
Let $D(h)$ be an irrep of $H$, and define $E(h)$ as
\begin{equation}
E(h) = \left[\frac{\omega(h,T_0)}{\omega(T_0,T_0^{-1}hT_0)}\right]^* D(T_0^{-1}hT_0),
\end{equation}
which also forms an irrep of $H$ but with a conjugated factor system relative to the original one.
Then, the standard forms for the three types are given by:

1. Type I:
\begin{equation}
U(h) = D(h), \quad U(T_0) = W;
\end{equation}

2. Type II:
\begin{equation}\label{type-II}
U(h) = \begin{pmatrix}
D(h) & 0 \\
0 & D(h)
\end{pmatrix}, \quad U(T_0) = \begin{pmatrix}
0 & -W \\
W & 0
\end{pmatrix};
\end{equation}

3. Type III:
\begin{equation}
U(h) = \begin{pmatrix}
D(h) & 0 \\
0 & E(h)^*
\end{pmatrix}, \quad U(T_0) = \begin{pmatrix}
0 & \omega(T_0, T_0)D(T_0^2) \\
I & 0
\end{pmatrix},
\end{equation}
where $ W $ is a unitary matrix satisfying $ WW^* = \pm \omega(T_0, T_0)D(T_0^2) $, with the plus sign applying to type I and the minus sign to type II.  Here we omit the anti-unitary operation $\kappa$ in the expression for $U(T_0)$. 
The complete proof can be found in the literature\cite{Shaw1974_2}. In the factor system $\omega$, the anti-unitary part does not introduce additional degeneracy for type I, unlike type II, which has additional degeneracy. For type II, the key condition is that $ WW^* = -\omega(T_0,T_0)D(T_0^2) $. One can prove that there exists a unique factor system $\omega'$ satisfying $\omega'(T_0,T_0) = -\omega(T_0,T_0)$, such that the type II irreducible co-representations for $\omega$ become type I for $\omega'$, as now $WW^* = +\omega'(T_0,T_0)D(T_0^2) $. Physically, the degeneracy in type II cases, determined by the sign $-1$, is analogous to Kramers' degeneracy. Specifically, the two factor systems $\omega'$ and $\omega$ differ by the non-trivial factor of the group $G/H \cong {Z}_2^{T}$, where $T^2=-1$. 
For type III, the degeneracy arises from induction, i.e., $ H \uparrow G $. This is why we can constrain the lower left corner of $ U(T_0) $ to be the identity matrix, which effectively identifies two different carrier spaces, simplifying the matrix form.

Upon determining the norm and $D(h)$ for the unitary part, we can use the previously mentioned equations to construct the standard form.

\subsection{diagonalize the representation matrices}

The procedure for diagonalizing the representation matrices also utilizes the Hamiltonian method, similar to the approach used for reducing representations. 
This method selects a normal subgroup $G_1$ and restrict the irrep of $G$ onto $G_1$, denoted as $U(g\in G)\downarrow G_1$.
The restricted representation may be reducible or irreducible. If reducible, a similarity transformation  $P$ can be found to diagonalize $U(g\in G_1)$.
However, this transformation often off-diagonalizes the representation matrices outside of $G_1$.

For co-irreps with norms of 2 or 4, we can initially choose $ G_1 = H $.
Through the reduction process $U(G)\downarrow H$, we obtain $P^{-1} U(h) P = E^*(h)\oplus D(h)$, where $D(h)$ is crucial for establishing the standard form of representation matrices, can be readily extracted from this expression.
The strategy involves iteratively selecting a unitary subgroup and reducing its representation until the subgroup's matrices are fully diagonalized. At this point, each matrix has only one non-zero element per row and column, completing the diagonalization process.

\subsection{solve the phase arbitrariness}

The remaining challenge in optimizing representation matrices lies in addressing the phase arbitrariness inherent in the anti-unitary component. 
For representations of dimension $d_r$, a similarity transformation
\begin{equation}
P=\text{diag}(e^{i\theta_1},e^{i\theta_2}, \ldots, e^{i\theta_{d_r}}),
\end{equation}
can be applied. This transformation preserves the matrix form of unitary elements while inducing a phase shift in the anti-unitary part.
The goal is to adjust as many matrix elements as possible to 1, thereby standardizing the representations.

The detailed steps for optimizing the representation matrices are outlined below:

\begin{enumerate}
\item Determine the norm of the representation. For norm 1, $D(h)$ directly represents the unitary segment.
For norms 2 or 4, decompose the restricted representation $U(g\in G)\downarrow H = C(h) \oplus D(h)$ and select one as $D(h)$ (for norm 4, both are linearly equivalent).
\item Iteratively select a unitary normal subgroup and check whether the restricted representation is reducible. If it is, apply a similarity transformation to diagonalize the matrices for this subgroup. Continue this process until the representation matrices of the selected subgroup are fully diagonalized.
\item Address the phase arbitrariness in the anti-unitary parts by adjusting as many matrix elements as possible to 1.
\item Construct the standard form from $D(h)$ according to the norm of the rep.
\end{enumerate}

It must be acknowledged that optimizing representation matrices requires significantly more effort compared to obtaining their initial forms. This process involves enumerating normal subgroups and performing multiple reductions. Thus, in most cases—such as when only the trace of the matrices is needed or for further computational use—the optimization step may be unnecessary. Even if the initial matrices appear more complex, they are strictly equivalent to their optimized forms. The described procedures merely find a similarity transformation for a more concise matrix form.

\section{$k\cdot p$ results of \ch{Mn3Sn}}\label{sm:kp matrices}
Regarding the 3D weyl point on the $\Gamma-A$ path,
we first obtained the reducible rep by taking the direct sum of the irreps  $^S \bar{DT}_3$ and $^S \bar{DT}_5$.
Following this, we applied the $k \cdot p$ method to this reducible rep. From this procedure, we derived three independent $k \cdot p$ Hamiltonians at the first order of $\bm{k}$, which are given by:
\begin{equation}
    \begin{aligned}
        H^{(1)}_k&=a\begin{pmatrix}
  1& 0 &0 \\
  0& 0 &0 \\
  0& 0 &0
\end{pmatrix}\delta k_z+b\begin{pmatrix}
  0& 0 &0 \\
  0& 1 &0 \\
  0& 0 &1
\end{pmatrix}\delta k_z\\
&+c \left [   \begin{pmatrix}
  0&\sqrt{3}+i  &-1-\sqrt{3}i \\
  \sqrt{3}-i &0  &0 \\
  -1+\sqrt{3}i&0  &0
\end{pmatrix}\delta k_x +
\begin{pmatrix}
  0&-\sqrt{3}+i  &-1+\sqrt{3}i \\
  -\sqrt{3}-i &0  &0 \\
  -1-\sqrt{3}i&0  &0
\end{pmatrix}\delta k_y \right ]
    \end{aligned}
\end{equation}

And the four independent $k \cdot p$ Hamiltonians at the second order of $\bm{k}$ deriving from SSG irrep $^S \bar{A}_3$:
\begin{equation}
    \begin{aligned}
        &H^{(2)}_k=a\begin{pmatrix}
  1&  0&0  &0 \\
  0&1  &0  &0 \\
  0& 0 & 1 &0 \\
  0&0  &0  &1
\end{pmatrix}(\delta k_z)^2\\&+b\left [
\begin{pmatrix}
  0&  0&-1-\sqrt{3}i   &0 \\
  0&0  &0  &-1+\sqrt{3}i \\
  -1+\sqrt{3}i& 0 & 0 &0 \\
  0&-1-\sqrt{3}i  &0  &0
\end{pmatrix}\delta k_x \delta k_z+
\begin{pmatrix}
  0&  0&-1+\sqrt{3}i   &0 \\
  0&0  &0  &-1-\sqrt{3}i \\
  -1-\sqrt{3}i& 0 & 0 &0 \\
  0&-1+\sqrt{3}i  &0  &0
\end{pmatrix}\delta k_y \delta k_z
\right ]\\
&+c\left [
\begin{pmatrix}
  0& \sqrt{3}-i &0  &0 \\
  \sqrt{3}+i& 0 &0  &0 \\
 0 & 0 & 0 &\sqrt{3}+i \\
 0 & 0 &\sqrt{3}-i  &0
\end{pmatrix}(\delta k_x)^2 +
\begin{pmatrix}
  0& 4i &0  &0 \\
  -4i& 0 &0  &0 \\
 0 & 0 & 0 &-4i \\
 0 & 0 &4i  &0
\end{pmatrix}\delta k_x \delta k_y -
\begin{pmatrix}
  0& \sqrt{3}+i &0  &0 \\
  \sqrt{3}-i& 0 &0  &0 \\
 0 & 0 & 0 &\sqrt{3}-i \\
 0 & 0 &\sqrt{3}+i  &0
\end{pmatrix}(\delta k_y)^2
\right ]\\
&+d 
\begin{pmatrix}
  1&  0&0  &0 \\
  0&1  &0  &0 \\
  0& 0 & 1 &0 \\
  0&0  &0  &1
\end{pmatrix}\left [
(\delta k_x)^2 + (\delta k_y)^2 - \delta k_x \delta k_y
\right ]
    \end{aligned}
\end{equation}

If we focus solely on the dispersion along the A-L path, where $\delta k_y = \delta k_z = 0$, only the $(\delta k_x)^2$ term remains. This is consistent with the $k \cdot p$ results discussed in the main text.

\nocite{*}

\end{document}